\documentclass[a4paper,10pt]{article}
\usepackage[a4paper, total={6.0in, 8in}]{geometry}
\usepackage[utf8]{inputenc}
\usepackage{color}
\usepackage{graphicx}

\newcommand{\be}{\begin{equation}}
\newcommand{\ee}{\end{equation}}
\newcommand{\bea}{\begin{eqnarray}}
\newcommand{\eea}{\end{eqnarray}}
\newcommand{\ba}[1]{\begin{array}{#1}}
\newcommand{\ea}{\end{array}}
\newcommand{\nn}{\nonumber}

\newcommand{\ep}{\epsilon}
\newcommand{\om}{\omega}  
\newcommand{\vk}{\vec k}

\newcommand{\del}{\partial}

\begin{document}

\title{From Non-interacting to Interacting Picture of Quark Gluon Plasma in presence of a magnetic field and its fluid property}

\author{Jayanta Dey$^1$, Sarthak Satapathy$^1$, Ankita Mishra$^2$, Souvik Paul$^3$,
\\
Sabyasachi Ghosh$^1$}
\date{}
\maketitle
\begin{center}
{$^1$Indian Institute of Technology Bhilai, GEC Campus, Sejbahar, Raipur 492015, Chhattisgarh, India}
\\
{$^2$ Department of Mechanical Engineering, Guru Ghasidas University, Bilaspur 495009, India}
\\
{$^3$ Department of Physical Sciences,Indian Institute of Science Education and Research Kolkata, 
Mohanpur, West Bengal 741246, India}
\end{center}

\begin{abstract}
We have attempted to build a parametric based simplified and analytical model to map the interaction of quarks and gluons
in presence of magnetic field, 
which has been constrained by quark condensate and thermodynamical quantities like pressure, energy density etc., obtained from the
calculation of lattice quantum chromodynamics. 
To fulfill that mapping,
we have assumed a parametric temperature and magnetic field dependent degeneracy factor, average energy, momentum and velocity of quarks and gluons. Implementing this QCD interaction in calculation of transport
coefficient at finite magnetic field, we have noticed that magnetic field and interaction both are 
two dominating sources, for which the values of transport coefficients can be reduced. Though the 
methodology is not so robust, but with the help of its simple parametric expressions, one can get 
a quick rough estimation of any phenomenological
quantity, influenced by temperature and magnetic field dependent QCD interaction.

\end{abstract}

\maketitle

\section{Introduction}
Extremely strong magnetic fields have been known to exist during the 
electroweak phase transition of the universe as suggested by cosmological models ~\cite{Vachaspati}.
Large values of magnetic fields are also present in the interior of dense neutron 
stars called magnetars ~\cite{Duncan}. 
Studying quantum field theory in the presence of magnetic field has led to many 
interesting observations such as magnetic catalysis ~\cite{Shovkovy}, chiral
magnetic effect ~\cite{Fukushima}, inverse magnetic catalysis ~\cite{Bali1,Bali2} and many more.
These phenomena might be indirectly observed in the laboratories of heavy ion collision (HIC) experiments
like RHIC and LHC, where an approximately $m_{\pi}^2 - 10m_{\pi}^2$ 
magnetic field is expected to be produced after collision
due to two opposite heavy-ionic currents~\cite{Tuchin:2013ie}.
Refs.~\cite{Deng:2012pc,Satow:2014lia,Skokov:Illarionov} have addressed a
possible space time evolution of electromagnetic fields, produced in the laboratories
of HIC experiments. Refs.~\cite{Roy:2015kma,Pu:2016ayh,Hongo:2013cqa,Inghirami:2016iru}
(also relevant references therein) have studied on evolution picture of expanding tiny 
medium in presence of magnetic field through the equations of magneto hydrodynamic (MHD). 
Impact of external magnetic field through
transport simulation can be noticed in Ref.~\cite{Das:2016cwd}.
Dissipative picture of MHD or transport simulation need a field dependent transport coefficients
inputs and therefore, a parallel microscopic
calculation of transport coefficients in present of magnetic field~\cite{Landau} is an important
topic in the community of heavy ion physics.
The transport coefficients in presence of magnetic field are recently investigated in
Refs.~\cite{Tuchin,Li_shear,Asutosh,JD_eta,NJLB_eta,Nam,HRGB,BAMPS,HM3,Denicol,
Sedrakian_el,Kerbikov:2014ofa,Nam:2012sg,Huang:2011dc,Hattori:2016lqx,
Manu1,Manu2,Feng_cond,Fukushima_cond,Arpan1,Arpan2,NJLB_el,SRath1,Lata,BC,Manu_SA,HRGB_QM,
Hattori_bulk,Sedarkian_bulk,Agasian_bulk1,Agasian_bulk2,Manu3,Manu4,Balbeer},
where field impact in shear viscosity~\cite{Tuchin,Li_shear,Asutosh,JD_eta,NJLB_eta,Nam,HRGB,BAMPS,HM3,Denicol}, 
electrical conductivity~\cite{Sedrakian_el,Kerbikov:2014ofa,Nam:2012sg,Huang:2011dc,Hattori:2016lqx,
Manu1,Manu2,Feng_cond,Fukushima_cond,Arpan1,Arpan2,NJLB_el,SRath1,Lata,BC,Manu_SA,HRGB_QM},
bulk viscosity~\cite{Hattori_bulk,Sedarkian_bulk,Agasian_bulk1,Agasian_bulk2,Manu3}
for light quark sector
as well as heavy quark sector~\cite{Manu4,Balbeer} are investigated.
Present article has gone through similar directional investigation.

With respect earlier works, present work has a unique combination of two components.
One is the mapping of QCD interaction at finite
temperature and magnetic field via fitting QCD thermodynamic, provided by recent lattice 
quantum chromodynamics (LQCD) calculations~\cite{Bali1,Bali2}. Another is  
a detail multi-component anatomy of shear viscosity and electrical conductivity
of quark gluon plasma in presence of magnetic field, which has been considered for that interacting picture.
Without field picture, we can find a long list
of Refs.~\cite{Gorenstein,Peshier,Heinz,Bluhm,Bannur,Salvatore,Meisinger,
Ruggieri,Chandra_PRC07,Chandra_EPJC09,Chandra_PRD11} and references therein, where
a temperature dependent QCD interaction is mapped but that much investigations are not found for
magnetic field picture. 
After the LQCD results of thermodynamics in presence of magnetic field~\cite{Bali1,Bali2},
a revised effort from effective QCD model calculation~\cite{PB1_19,PB2_Tawfik,PB3_Tawfik,PB4_Farias}
have been attempted. In this context, present work has attempted to find a simple
parametric mapping of LQCD thermodynamics in presence of magnetic field~\cite{Bali1,Bali2}.

In earlier Ref.~\cite{SS_JPG} by Satapathy et al., an attempt is made for building a simplified parametric model
through matching the LQCD thermodynamics without magnetic field. There, a temperature dependent
degeneracy factor of QGP system is prescribed to map the temperature dependent QCD interaction. 
Extending the earlier study of Ref.~\cite{SS_JPG}, a quasi-particle model at 
finite magnetic field picture is attempted in present article, where a temperature
and magnetic field dependent degeneracy factor is first proposed by matching the LQCD thermodynamics 
in presence of magnetic field~\cite{Bali1,Bali2}, then they are used for estimating anisotropic
components of transport coefficients of QGP. 

The article is organized as follows. Next in Sec.~(\ref{sec:th_TB}), the mapping of LQCD
thermodynamics in presence of magnetic field has been addressed.
After developing the quasi particle description, it is applied to estimate transport
coefficients like shear viscosity and electrical conductivity at finite magnetic 
field in Sec.~(\ref{sec:sh_el}), whose framework is briefly addressed in Sec.~(\ref{sec:App}) with
two subsections - (\ref{sec:Sh_B}) and (\ref{sec:el_B}). At the end in Sec.~(\ref{sec:sum})
we have summarized our investigations.

\section{Mapping LQCD Thermodynamics at finite magnetic field}
\label{sec:th_TB}
%
Owing to the basic statistical mechanics, pressure $P$, number density $n$ and energy density $\epsilon$ of grand canonical ensemble at finite temperature $T=1/\beta$ and zero chemical potential ($\mu=0$) can be obtained from its partition function $Z$ and their final expressions are respectively
\bea
P &=&\frac{T}{V} \ln Z 
\nn\\
&=& g\frac{T}{a}\int \frac{d^3k}{(2\pi)^3}\ln \{1+ae^{-\beta E}\}
\nn\\
&=&g\int \frac{d^3k}{(2\pi)^3}\Big[\frac{k^2}{3E}\Big]\frac{1}{e^{\beta E}+a}~,
\\
n &=& g\int \frac{d^3k}{(2\pi)^3}\Big[1\Big]\frac{1}{e^{\beta E}+a}~,
\\
\epsilon &=& g\int \frac{d^3k}{(2\pi)^3}\Big[E\Big]\frac{1}{e^{\beta E}+a}~,
\label{GCE}
\eea
where $g$ is degeneracy factor, and $a=\pm 1$ for Fermion and Boson medium. 

Using above equations for massless quark gluon plasma (QGP) system, we will get:
\bea
P_{QGP}&=&g_g\int \frac{d^3k}{(2\pi)^3}\Big[\frac{k^2}{3E}\Big]\frac{1}{e^{\beta E}-1}+g_Q\int \frac{d^3k}{(2\pi)^3}\Big[\frac{k^2}{3E}\Big]\frac{1}{e^{\beta E}+1}
\nn\\
&=&\Big[g_g+g_Q\Big(\frac{7}{8}\Big)\Big]\frac{\zeta(4)}{\pi^2}T^4
=\Big[g_g+g_Q\Big(\frac{7}{8}\Big)\Big]\frac{\pi^2}{90}T^4\approx 5.2~T^4~,
\\
n_{QGP}&=&g_g\int \frac{d^3k}{(2\pi)^3}\Big[1\Big]\frac{1}{e^{\beta E}-1}+g_Q\int \frac{d^3k}{(2\pi)^3}\Big[1\Big]\frac{1}{e^{\beta E}+1}
\nn\\
&=&n_g + n_Q=\Big[g_g+g_Q\Big(\frac{3}{4}\Big)\Big]\frac{\zeta(3)}{\pi^2}T^3\approx 5.23~T^3~,
\\
\epsilon_{QGP}&=&g_g\int \frac{d^3k}{(2\pi)^3}\Big[E\Big]\frac{1}{e^{\beta E}-1}+g_Q\int \frac{d^3k}{(2\pi)^3}\Big[E\Big]\frac{1}{e^{\beta E}+1}
\nn\\
&=& \Big[g_g+g_Q\Big(\frac{7}{8}\Big)\Big]\frac{3\zeta(4)}{\pi^2}T^4
=\Big[g_g+g_Q\Big(\frac{7}{8}\Big)\Big]\frac{3\pi^2}{90}T^4\approx 15.6~T^4~,
\label{SB_QGP}
\eea
where $g_g=16$ and $g_Q=36$ are degeneracy factors of gluons and 3-flavor quarks. Here $n_{QGP}$ stands for total number density of QGP, which is different from net quark density (which is zero at zero quark chemical potential).   
Massless values of entropy density $s_{QGP}$ can also obtained as
\be
s_{QGP}=\frac{\epsilon +P}{T}=\frac{4P}{T}\approx 20.8 T^3~.
\ee
If we calculate average energy or momentum of massless Boson and Fermion medium, then we will get 
\bea
E^{g,Q}_{\rm av}(m=0)&=&k^{g,Q}_{\rm av}(m=0)=\frac{\int \frac{d^3p}{(2\pi)^3}\frac{E}{e^{\beta E}+a}}{\int \frac{d^3p}{(2\pi)^3}\frac{1}{e^{\beta E}+a}}
\nn\\
&=&3T\frac{\zeta(4)}{\zeta(3)}~{\rm for~gluon~or,~}a=-1~,
\nn\\
&=&\frac{7T}{2}\frac{\zeta(4)}{\zeta(3)}~{\rm for~quark~or,~}a=+1~,
\nn\\
&=&3T~{\rm for~spinless~parton~or,~}a=0~.
\label{E_m0}
\eea
Using those average values, one can rewrite $P_{QGP}$, $\epsilon_{QGP}$ in terms of $n_{QGP}$ as:
\bea
P_{QGP} &=& \Big[\frac{k_{\rm av}^2(m=0)}{3E_{\rm av}(m=0)}\Big]_g n_g+\Big[\frac{k_{\rm av}^2(m=0)}{3E_{\rm av}(m=0)}\Big]_Q n_Q
\nn\\
&=& \Big[\frac{k_{\rm av}(m=0)}{3}\Big]_g n_g+\Big[\frac{k_{\rm av}(m=0)}{3}\Big]_Q n_Q
\nn\\
&=&\Big[T\frac{\zeta(4)}{\zeta(3)}\Big] g_g\frac{\zeta(3)}{\pi^2}T^3+\Big[\frac{7T}{6}\frac{\zeta(4)}{\zeta(3)}\Big]g_Q\Big(\frac{3}{4}\Big)\frac{\zeta(3)}{\pi^2}T^3
\nn\\
&=&\Big[g_g+g_Q\Big(\frac{7}{8}\Big)\Big]\frac{\pi^2}{90}T^4\approx 5.2~T^4~,
\label{P_n}
\eea
\bea
\epsilon_{QGP} &=& \Big[E_{\rm av}(m=0)\Big]_g n_g + \Big[E_{\rm av}(m=0)\Big]_Q n_Q~,
\nn\\
&=&\Big[3T\frac{\zeta(4)}{\zeta(3)}\Big] g_g\frac{\zeta(3)}{\pi^2}T^3+\Big[\frac{7T}{2}\frac{\zeta(4)}{\zeta(3)}\Big]g_Q\Big(\frac{3}{4}\Big)\frac{\zeta(3)}{\pi^2}T^3
\nn\\
&=&\Big[g_g+g_Q\Big(\frac{7}{8}\Big)\Big]\frac{3\pi^2}{90}T^4\approx 15.6~T^4~.
\label{Pe_n}
\eea
In more simplified way, by considering average energy of spinless parton (a=0), we can write:
\bea
P_{QGP} &=& \Big[\frac{k_{\rm av}(m=0)}{3}\Big]\times n_{QGP}=\Big[\frac{3T}{3}\Big]\times 5.23T^3\approx 5.23 T^4
\nn\\
\epsilon_{QGP} &=& \Big[E_{\rm av}(m=0)\Big]\times n_{QGP}=\Big[3T\Big]\times 5.23T^3\approx 15.69 T^4~.
\label{QGP_MB}
\eea
Reader may notice very negligeable difference between Eq.~(\ref{QGP_MB}) and Eqs.~(\ref{P_n}), (\ref{Pe_n}), so considering average energy and momentum of massless and spinless parton as $3T$ might not be too bad assumption to consider.

\begin{figure}
 \centering
 \includegraphics[scale=0.3]{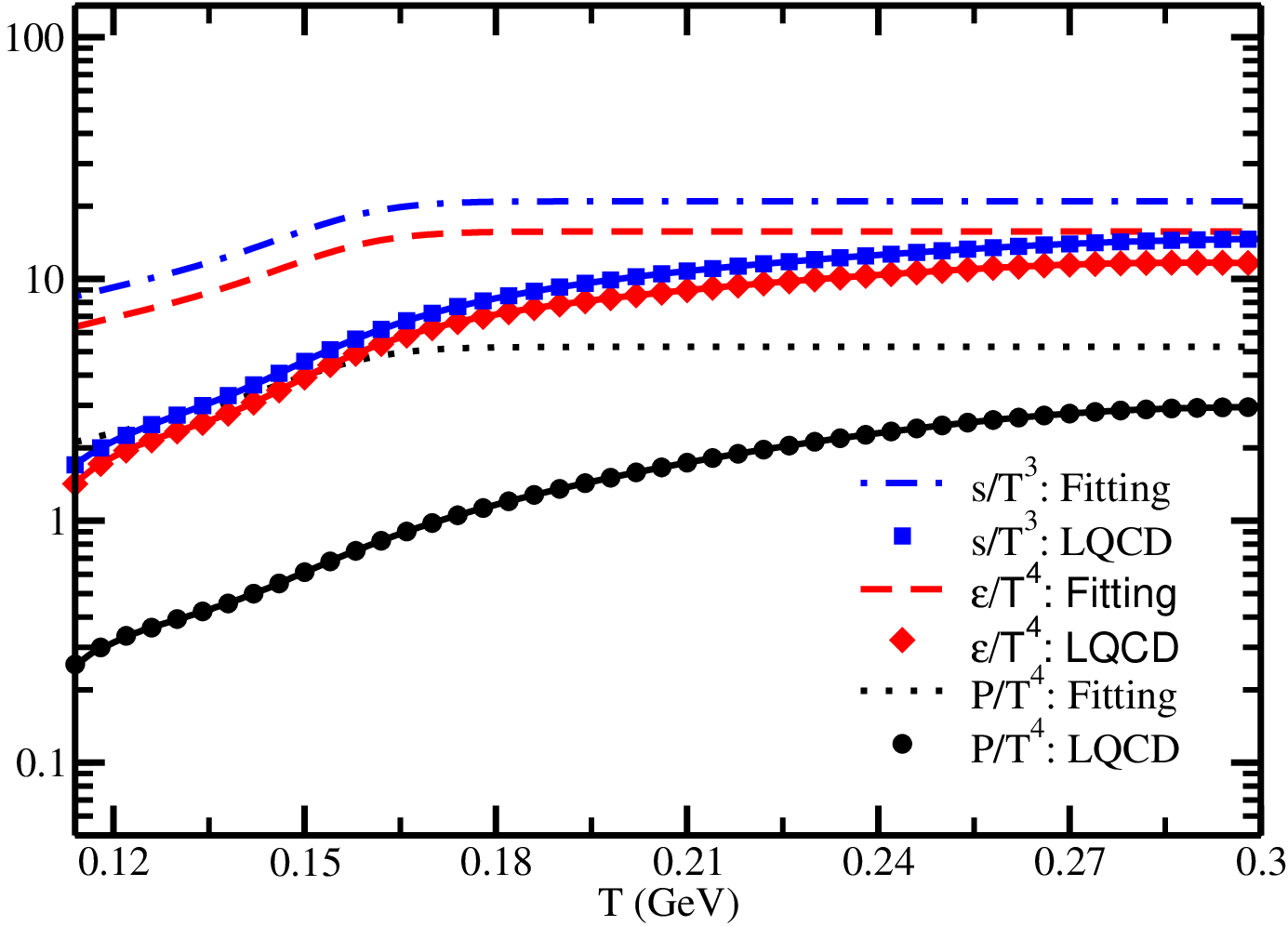}  
 \includegraphics[scale=0.3]{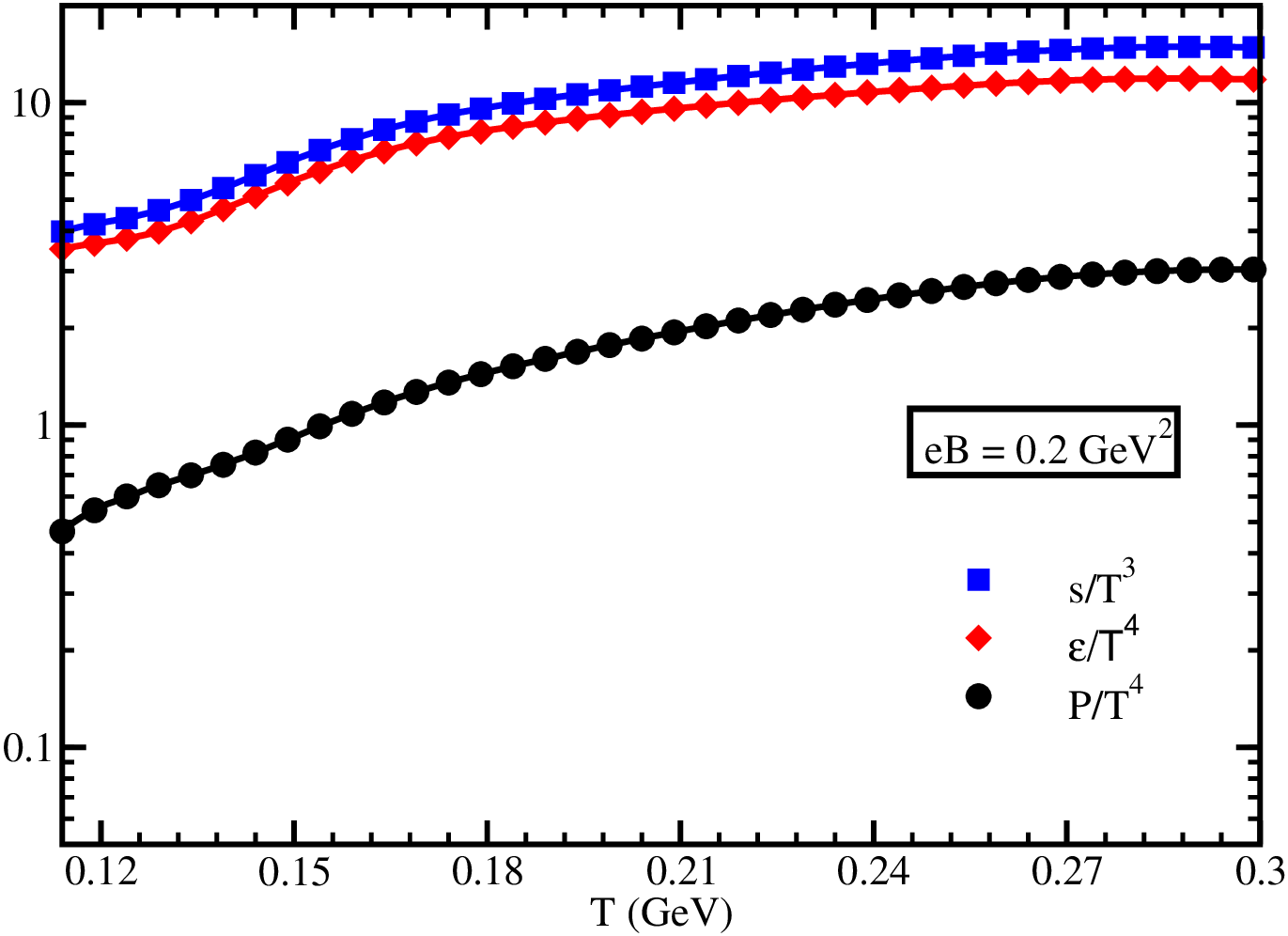} 
 \includegraphics[scale=0.3]{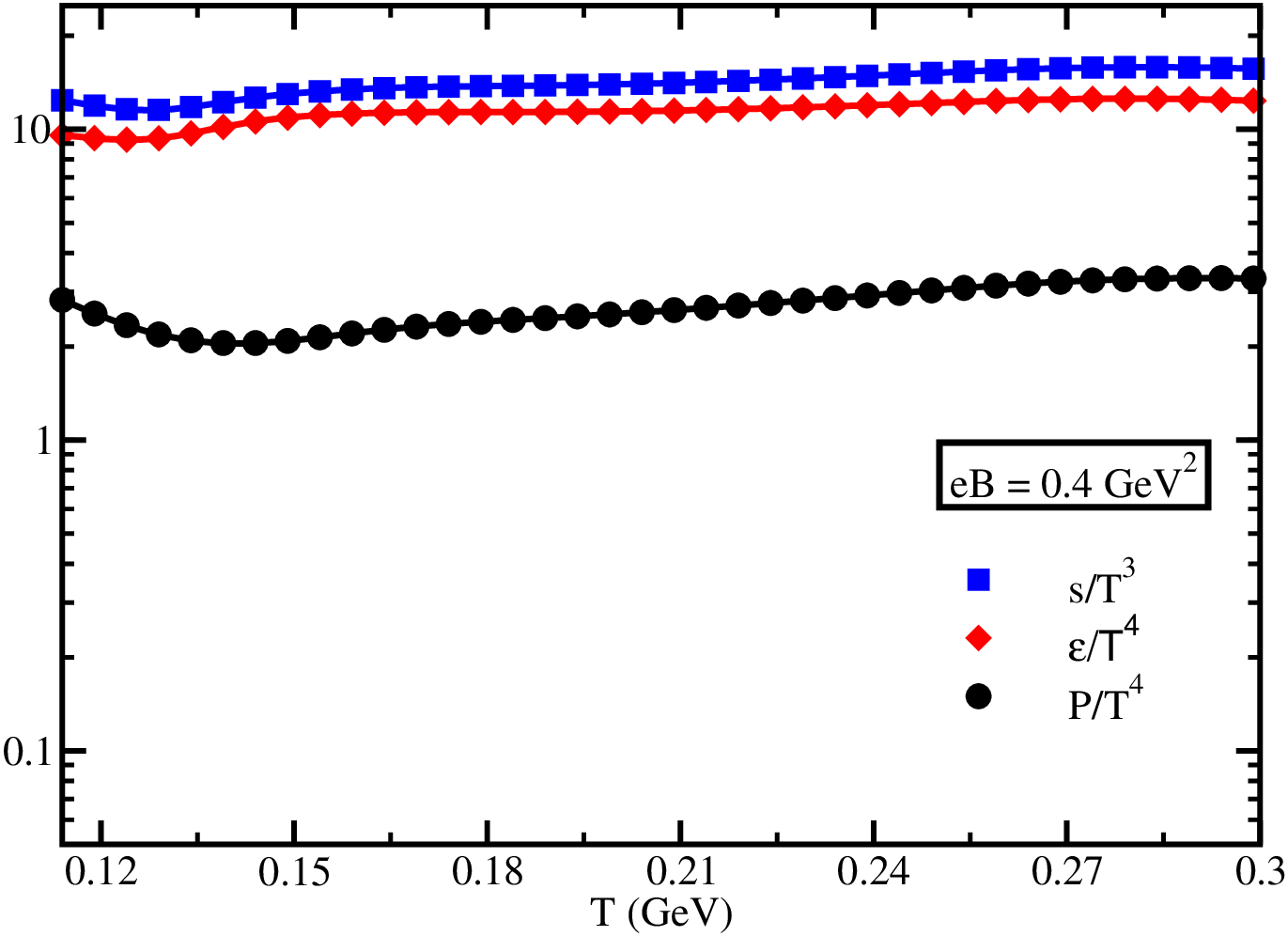} 
 \includegraphics[scale=0.3]{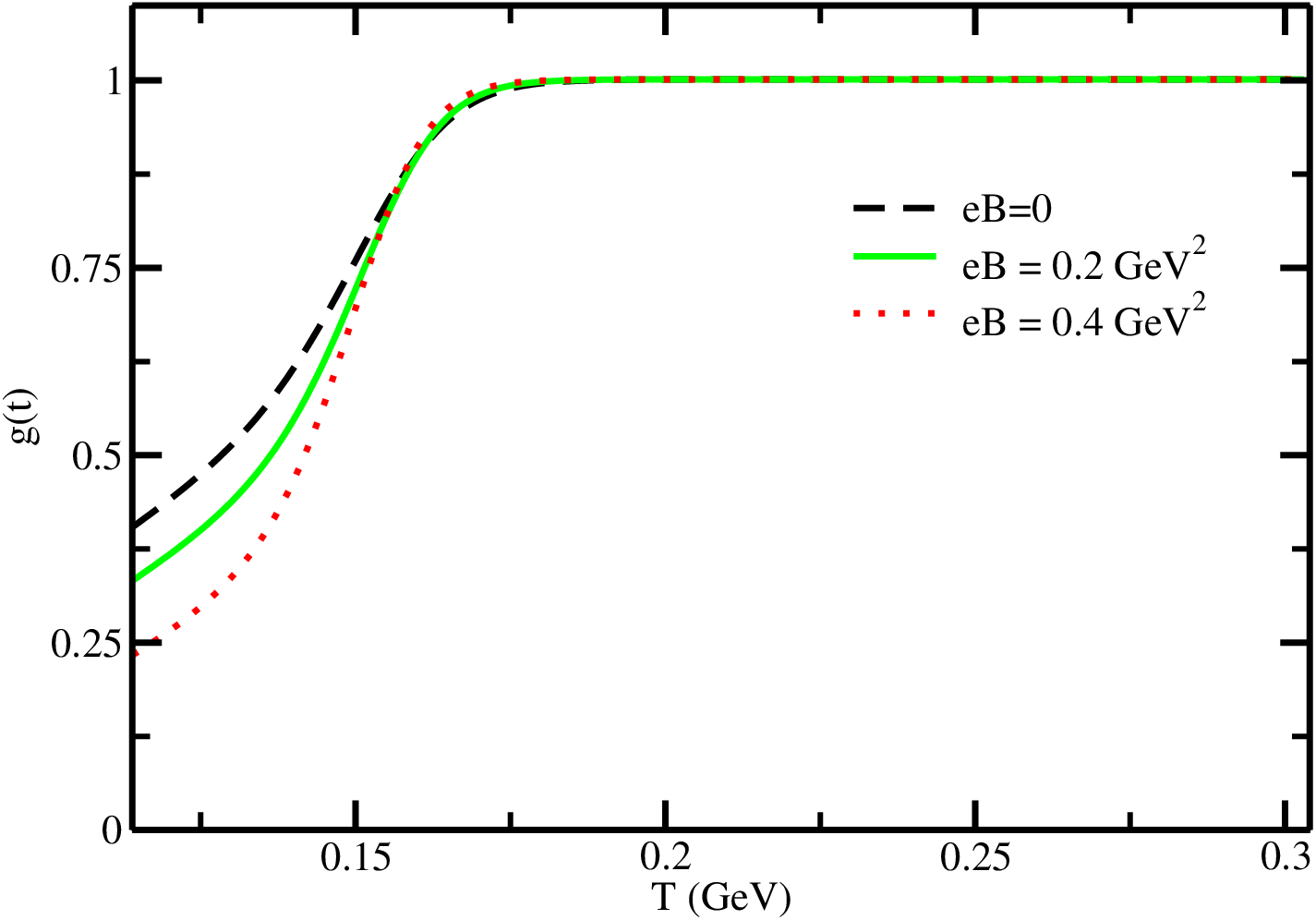}
 \includegraphics[scale=0.3]{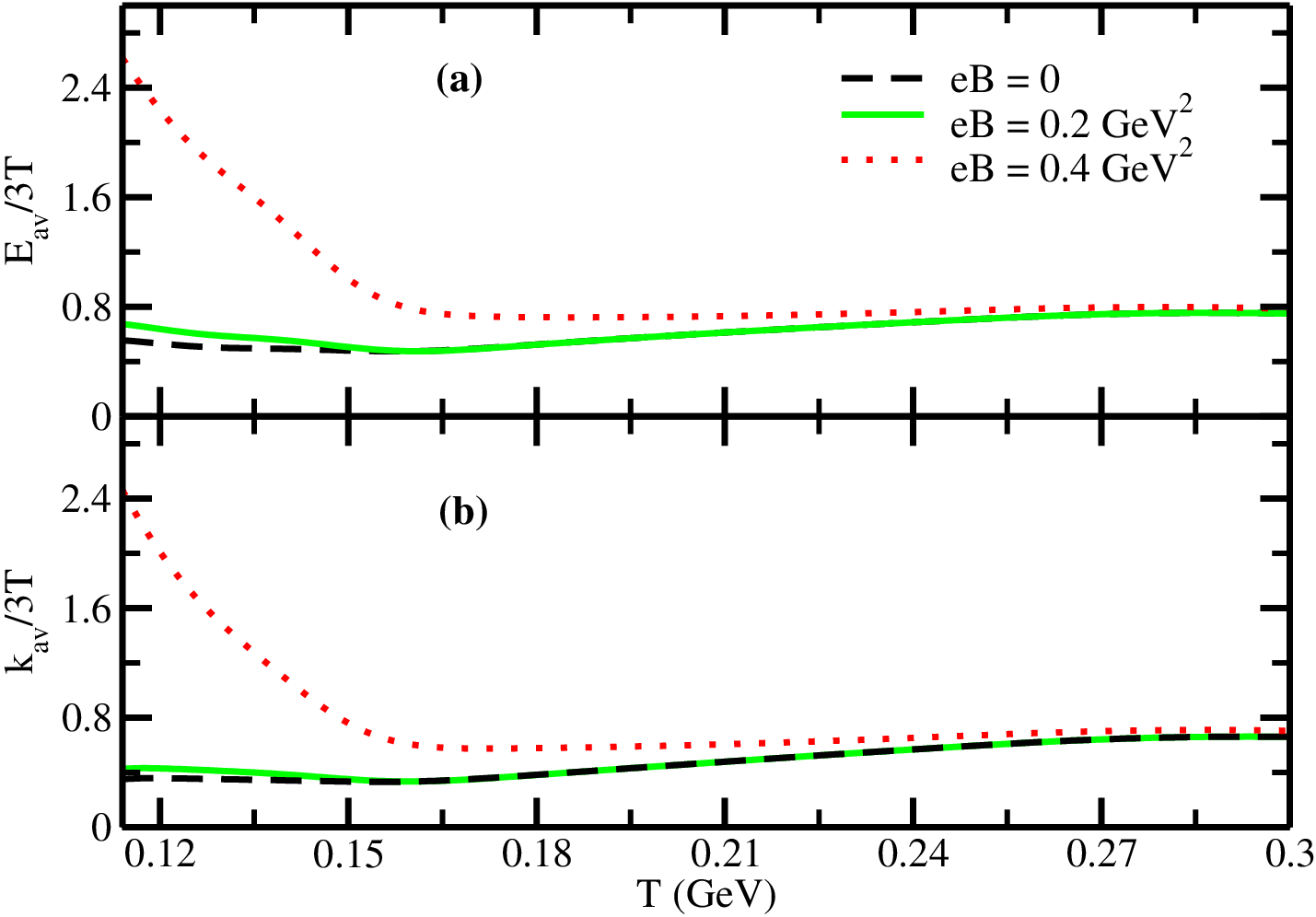}
 \includegraphics[scale=0.3]{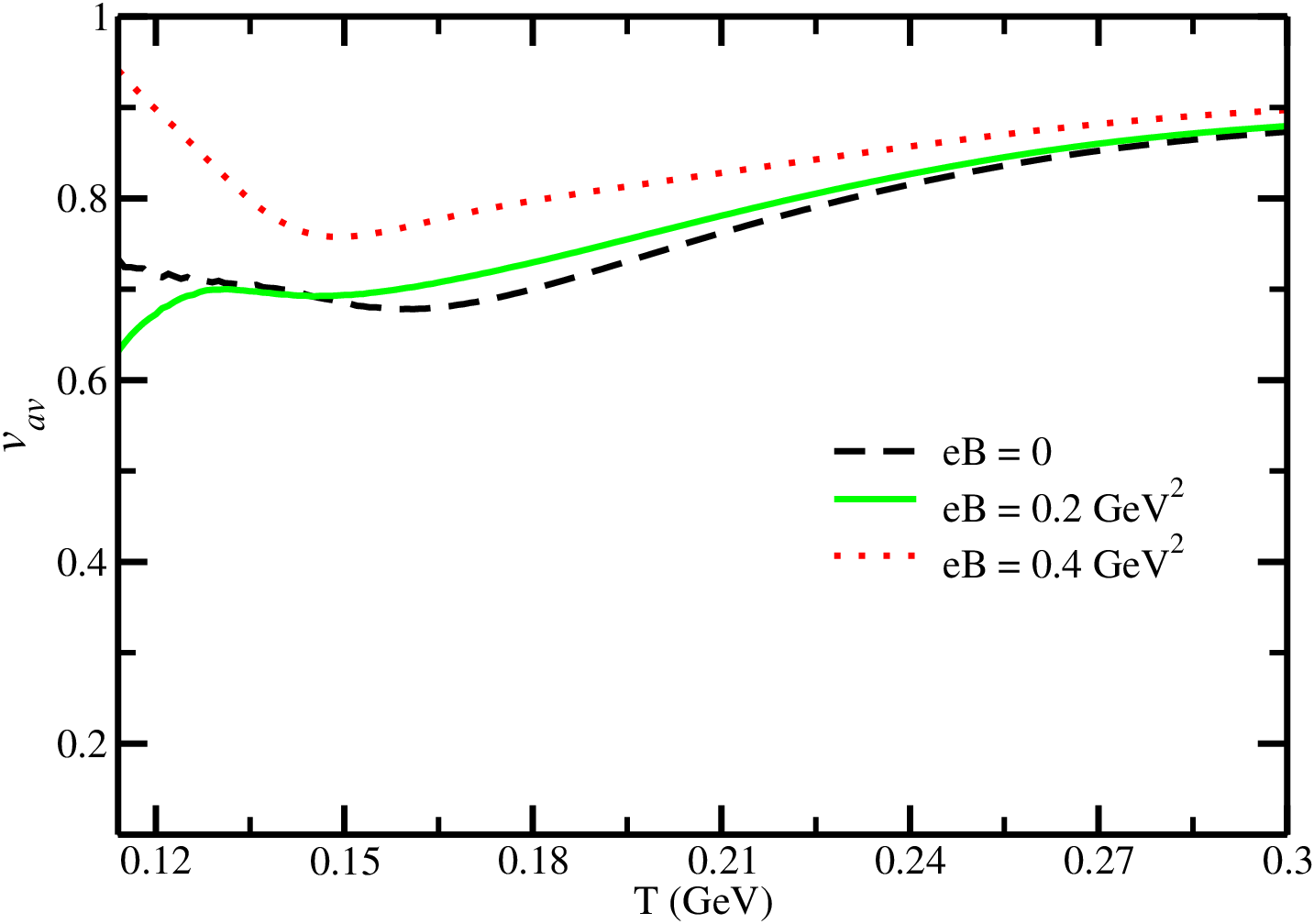}
 \caption{LQCD data of pressure (black circles), energy density (red diamonds) and entropy density (blue squares) are fitted at $eB=0$ (Left-upper panel), $eB=0.2$ GeV$^2$ (Right-upper panel), $eB=0.4$ GeV$^2$ (Left-middle panel) by our proposed model (solid curves) with tuning parameters degeneracy factor $g(T,B)$ (Right-middle panel), average energy $E_{\rm av}(T,B)$, momentum $k_{\rm av}(T,B)$ (Left-lower panel) and velocity $v_{\rm av}(T,B)$ (Right-lower panel). Using only $g(T)$ tuning parameter, dotted, dash and dash-dotted lines in left-upper panel are showing the deviation from LQCD data points, which is resolved by other tuning parameters.}
 \label{gvsT}
 \end{figure}

High $T$ quantum chromo dynamics (QCD) matter might behave like this and these massless values are popularly known as Stefan Boltzmann (SB) limits. It is lattice QCD (LQCD), who provide us
a better picture thermodynamical quantities like
$P$, $\epsilon$, $s$ etc. As we go from high to low $T$ range, a reduced values of thermodynamical quantities with respect their SB limits are observed in LQCD picture.
There are different quasi-particle type frameworks
~\cite{Gorenstein,Peshier,Heinz,Bluhm,Bannur,Salvatore,Meisinger,
Ruggieri,Chandra_PRC07,Chandra_EPJC09,Chandra_PRD11} (see also references therein) have been attempted to map the reduced values of LQCD thermodynamics. In this context, we have tried to map this fact as a reduction of degeneracy factors. If we go below quark-hadron transition temperature ($T_c$), then we will get hadronic matter (HM), where hadrons are the relevant degrees of freedom
of the system. If we assume that pion and kaon as most abundant mesons, made by u, d and s quarks, then for a quick estimations, one can calculate their massless expressions of thermodynamical quantities:
\bea
P_{HM}&=&(g_\pi +g_K)\int \frac{d^3k}{(2\pi)^3}\Big[\frac{k^2}{3E}\Big]\frac{1}{e^{\beta E}-1}
=(g_\pi +g_K)\frac{\zeta(4)}{\pi^2}T^4\approx 0.76~T^4~,
\nn\\
n_{HM}&=&(g_\pi +g_K)\int \frac{d^3k}{(2\pi)^3}\Big[1\Big]\frac{1}{e^{\beta E}-1}
=(g_\pi +g_K)\frac{\zeta(3)}{\pi^2}T^3\approx 0.77~T^3~,
\nn\\
\ep_{HM}&=&(g_\pi +g_K)\int \frac{d^3k}{(2\pi)^3}\Big[E\Big]\frac{1}{e^{\beta E}-1}
=(g_\pi +g_K)\frac{3\pi^2}{90}T^4\approx 2.28~T^4~,
\nn\\
s_{HM}&=&\frac{\epsilon_{HM} +P_{HM}}{T}
=(g_\pi +g_K)\frac{4\zeta(4)}{\pi^2}T^3\approx 3.04~T^3~,
\eea
where $g_\pi=3$, $g_K=4$ are degeneracy factors of $\pi$ and $K$ mesons respectively.
Now if we see the Lattice quantum chromodynamics (LQCD) data of $P(T)$, $\epsilon(T)$, $s(T)$, 
which is coppied from Ref.~\cite{Bali2} and pasted in Fig.~\ref{gvsT}(a), then one can
notice that the data points are located within the ranges $0.76 <\frac{P}{T^4}<5.2$,
$2.28< \frac{\epsilon}{T^4}<15.6$, $3.04<\frac{s}{T^4}<20.8$. 
In this context, quark-hadron phase transition may be considered as transition between two massless values of thermodynamical quantities for QGP to HM system, which can
be realized as reduction of degeneracy factor from 
$\Big[g_g+g_Q\Big(\frac{7}{8}\Big)\Big]\approx 47.5$ to $(g_\pi +g_K)\approx 7$.
Hence, one can understand the smooth cross-over transition from QGP to HM phase through
smooth reduction of degeneracy factors of QGP system.
To execute this idea, temperature dependent fraction/factor $g(T)$ has to be first multiplied with thermodynamical 
quantities, given in Eq.~(\ref{SB_QGP}) and then by matching LQCD data,
one can get the parametric form of $g(T)$. 
Hence, $g(T)=1$ will correspond to non-interacting picture or (roughly) massless case or SB limits
of QGP, given in Eqs.~(\ref{SB_QGP}). While we will get interacting QGP when we use parametric form of $g(T)$, which will be
always less than one in entire temperature range. Refs.~\cite{Bali1,Bali2} have provided $P_{LQCD}$ and $\epsilon_{LQCD}$ data but not $n_{LQCD}$ data. Using their quark condensate ($\langle{\bar q}q\rangle_T$) data, we can go with rough estimate of constitute quark mass as $M\approx\frac{M_N}{3}\langle{\bar q}q\rangle_T$, with nucleon mass $M_N=0.940$ GeV. 
 \begin{figure}
 \centering
 \includegraphics[scale=0.24]{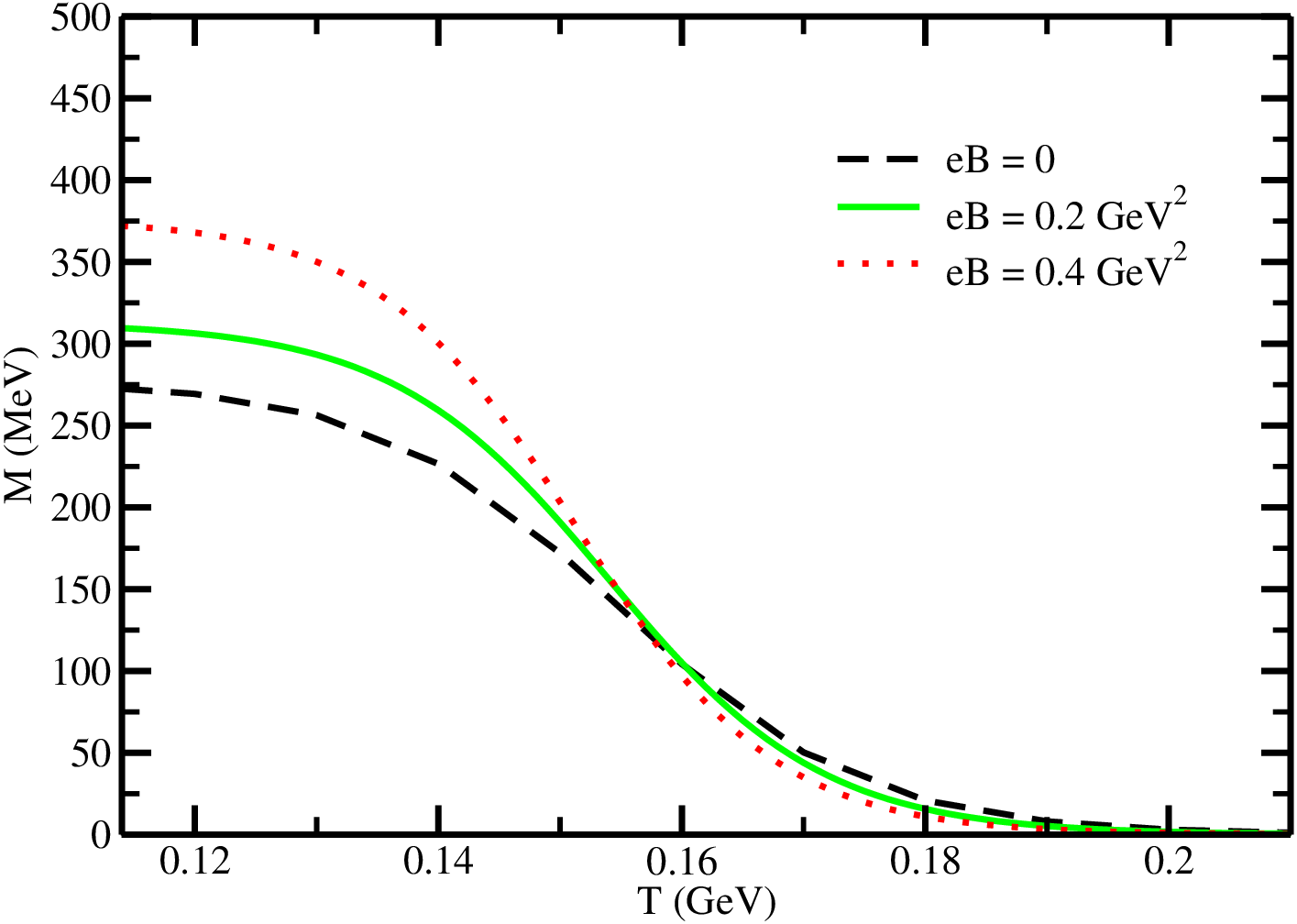} 
 \includegraphics[scale=0.24]{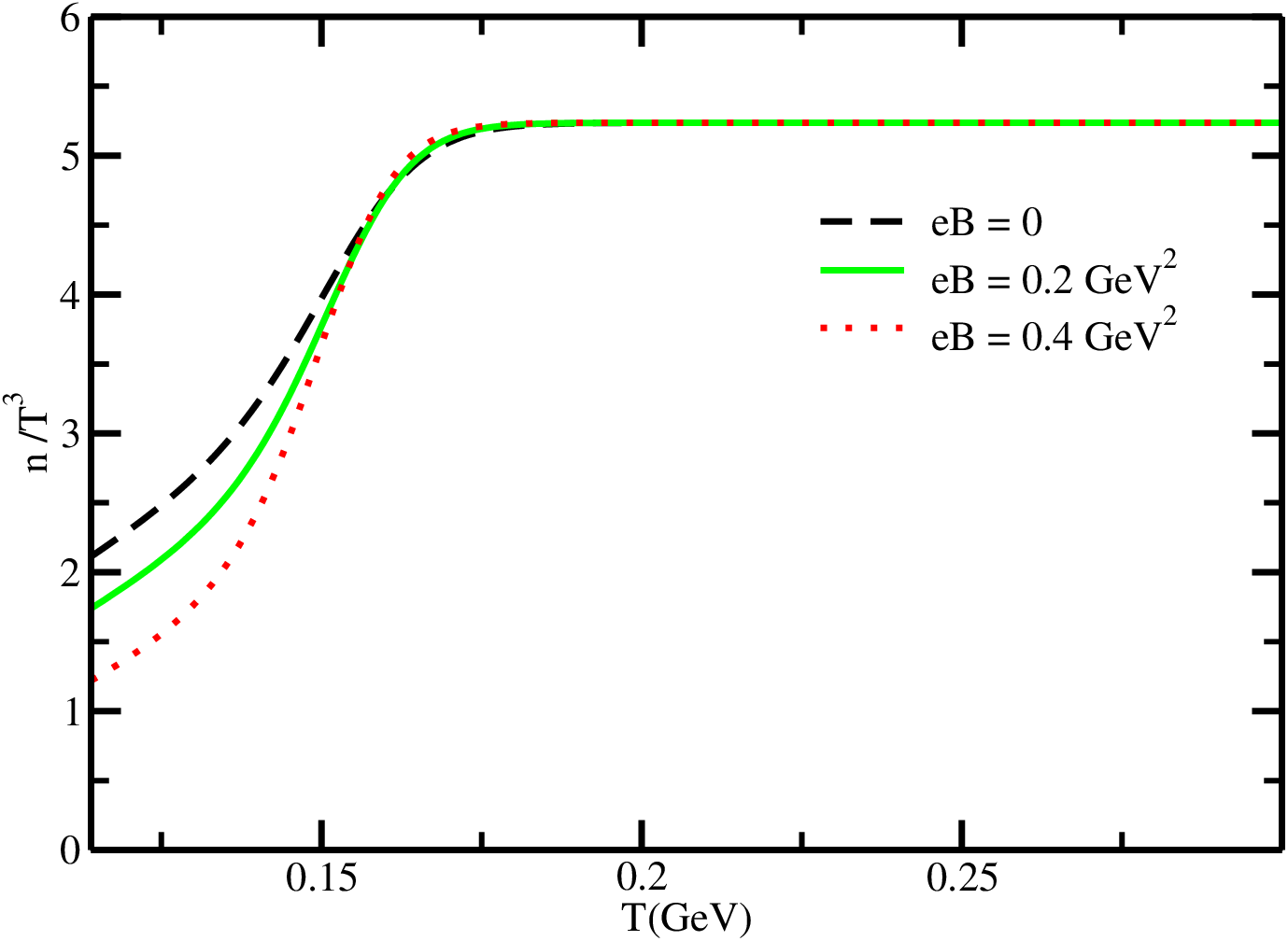} 
 \caption{Constituent quark mass $M$ (left) and normalized total number density $n_{LQCD}/T^3$ vs temperature $T$ at different magnetic field. The estimations are based on LQCD quark condensate data~\cite{Bali1}.}
 \label{M_n_T}
 \end{figure}
Reader can recognize the idea is guided from gap equation of NJL model with zero current quark mass. Then using that $M(T)$, we can estimate total number density. Let us called it as $n_{LQCD}$ and can be estimated by using the expressions:
\be
n_{LQCD}=g_g\int \frac{d^3k}{(2\pi)^3}\frac{1}{e^{\beta k}-1}+g_Q\int \frac{d^3k}{(2\pi)^3}\frac{1}{e^{\beta \sqrt{\vk^2+M^2(T)}}+1}~.
\ee
The $M(T)$ (left) and $n_{LQCD}(T)$ (right) are shown by black dash line in Fig.~(\ref{M_n_T}).
Now we have searched $g(T)$, which satisfy 
\be
n_{LQCD}=g(T)\times n_{QGP}=g(T)\times 5.23 T^3~.
\ee
Using that $g(T)$, we can get an rough estimation of pressure, energy density and entropy density as $g(T)\times P_{QGP}=g(T)\times 5.2 T^4$, $g(T)\times \epsilon_{QGP}=g(T)\times 15.6 T^4$ and $g(T)\times s_{QGP}=g(T)\times 20.8 T^3$ respectively, whose qualitative reduction trend is found as expected but they are quite far from the actual LQCD data values - $P_{LQCD}$,
$\epsilon_{LQCD}$ and $s_{LQCD}$. The dotted, dash and dash-dotted lines in left-upper panel of Fig.~(\ref{gvsT}) has explored this fact. We notice that they are far from LQCD data, given by circles, dimonds and squares.   

Now, matching all LQCD thermodynamical quantities through a single parameter
tuning is probably an impossible task, therefore more than one parameter can be helpful.
So along with $g(T)$, if we consider $E_{\rm av}$,
$k_{\rm av}$ as other tuning parameters, which might be deviated from massless expressions, given in Eq.~(\ref{E_m0}), we may get better fitted picture. 
Similar to the $T$ dependent fraction/factor $g(T)$, we can assume $f_E(T)$ and $f_k(T)$ can be multiplied with massless and spinless values of average energy and momentum as
\bea
E_{\rm av}(T) &=& 3T\times f_E(T)
\nn\\
k_{\rm av}(T) &=& 3T\times f_k(T)~,
\eea
where using standard relativistic relation among momentum, velocity and energy -
\be
k_{\rm av}(T) = v_{\rm av}(T)\times E_{\rm av}(T)~,
\ee
we can get a connection between average velocity $v_{\rm av}(T)$, $f_E(T)$ and $f_k(T)$ as
\be
f_k(T)=f_E(T)\times v_{\rm av}(T)~.
\ee
One can get $f_E=1$, $f_k=1$ and $v_{\rm av}(T)=1$ for massless case.
Now we can tune our $f_E(T)$ and $v_{\rm av}(T)$ by fitting LQCD data of $P_{LQCD}$ and $\epsilon_{LQCD}$ by imposing simplified quasi-particle relations:
\bea
\epsilon_{LQCD} &=& \Big[E_{\rm av}\Big] n_{LQCD}
\nn\\
&\approx& f_E(T)\times g(T)\times 15.69 T^4 
\nn\\
P_{LQCD} &=& \Big[\frac{k_{\rm av}(T) v_{\rm av}(T)}{3}\Big] n_{LQCD}
\nn\\
&\approx& v^2_{\rm av}(T)\times f_E(T)\times g(T)\times 5.23 T^4
\eea
Let us try to understand an overall idea of our proposed tuning set up. First we have attempted to map $T$-dependent quark condensate, which is in general mapped by constituent quark mass in effective QCD models but here we did it in different way. We first build number density data by using $T$ dependent quark mass, then imposing that still the total number density is basically number density of massless QGP, where their degeneracy factor mainly modified. So indirectly degeneracy factor carry the information of $T$-dependent quark condensate instead of constituent quark mass. Since number density is simply integration of thermal distribution, so we have use it as reference quantity to extract the temperature profile of tuning parameter - degeneracy factor. Just by multiplying $T$-dependent degeneracy factor with number density of massless QGP, one may get its corresponding values in interacting picture. Now for other thermodynamical quantities like pressure, energy density, which are roughly average values of energy and $\frac{1}{3}\times$momentum$\times$velocity, multiplied by number density.
Here we are assuming those average quantities, related with one-body kinematics will also be deviated from its massless limits as it happens for degeneracy factors. 
Hence by tuning the degeneracy factor first we are able to map quark condensate information of LQCD and next by tuning the average kinematics of spinless partons, we can manage to map the LQCD thermodynamics. 

 In presence of magnetic field along z-axis, we can get an anisotropy in pressure but if we consider pressure along z-direction ($P_z$), then thermodynamics relation
 \be
 s=\frac{\epsilon +P_z}{T}
 \ee
remain same~\cite{Bali1}. So, earlier expressions can be used for finite $B$ picture by tuning $T$, $B$ dependent parameters $g(T,B)$, $f_E(T,B)$, $f_k(T,B)$ or $v_{\rm av}(T,B)$~. Hence, $T$ and $B$ dependent LQCD thermodynamics~\cite{Bali1,Bali2} can be realized via simple parametric relations:
\bea
n_{LQCD}(T,B)&=&g(T,B)\times 5.23T^3
\nn\\
\epsilon_{LQCD}(T,B)&=&E_{\rm av}(T,V)\times n_{LQCD}(T,B)
\nn\\
&=&f_E(T,B)\times g(T,B)\times 15.69 T^4
\nn\\
P_{LQCD}(T,B)&=&\frac{v^2_{\rm av}(T,B)\times E_{\rm av}(T,B)}{3}\times n_{LQCD}(T,B)
\nn\\
&=&v^2_{\rm av}(T,B)\times f_E(T,B)\times g(T,B)\times 5.23 T^4~.
\eea
The fitted curves of LQCD data and the $T$, $B$ dependent tuning parameters $g(T,B)$, $E_{\rm av}(T,B)$, $k_{\rm av}(T,B)$ and $v_{\rm av}(T,B)$ are shown in Figs.~(\ref{gvsT}).

At the end of this section, let us briefly touch the well known inverse magnetic catalysis (IMC) effect, which was the key points of recent LQCD data~\cite{Bali1,Bali2}. The fact is as follows. The chiral condensate $\langle{\bar q}q\rangle_T$ in absence of magnetic field melts down near transition temperature $T_c$, when one goes from hadronic to quark temperature domain. In presence of magnetic field enhancement of condensate was very well known phenomena and known as magnetic catalysis (MC). LQCD data~\cite{Bali1,Bali2} at low $T$ show this MC but near transition temperature an IMC effect is noticed. Constituent quark mass $M(T,B)$, as shown in left panel of Fig.~(\ref{M_n_T}), is revealing that fact as it is proportionally mapped from condensate $\langle{\bar q}q\rangle_T(T,B)$. Further temperature derivative of mass or condensate we can get peak at $T_c$, which will be shifted towards lower values as we increase $B$. This reduction of $T_c$ with $B$ become well known graphical representation of IMC phenomena. However, in thermodynamical quantities like pressure, energy density, entropy density and transport coefficients like shear viscosity and electrical conductivity, this IMC fact becomes faint due thermal distribution ($\sim e^{-\beta M(T,B)}$). So searching effect of IMC might be tedious job from those quantities (addressed in present article). However, interaction measure $\epsilon - 3P$ and its connected transport coefficients like bulk viscosity (as $\zeta\propto(\epsilon - 3P)$) might able to expose the IMC pattern, where their peak might be shifted with $eB$.  
%
%
%
%
\subsection{Picture of reducing degeneracy factor}
To visualize the possibility of reduction degeneracy factor with decreasing temperature, let us recapitulate the similar fact of reducing degrees of freedom in di-atomic or n-atomic molecular system. At low temperature, di-atomic or n-atomic molecules have $3\times 2-1$ or $3\times n - k$ degrees of freedoms, because it carries 1 or k
number of atomic bindings. This bonding can be broken at high temperature and its degrees of freedom can be increased from $3\times 2-1=5$ to $3\times 2=6$ or from $3\times n - k$ to $3\times n$.
Based on the equipartition theorem, (internal) energy density of 
di-atomic or n-atomic molecular system will be proportional to its degrees 
of freedom, therefore internal energy or other thermodynamical quantities like pressure, entropy will
be enhanced with increasing temperature.
Similarly, when we go from hadronic phase (assuming abundant $\pi+K$ mesons) to QGP (u, d, s quarks and gluons) by increasing temperature, the degeneracy factor probably transform smoothly from $g=7$ to $g=52$. 

Formulating actual mechanism of this fact might be very difficult (but certainly a good problem of QCD sector), but it can be visualize by counting quantum states of
any hadron and its constituent's quantum states.
For example, a colorless $\pi^+$ state in the low T or hadronic phase can be melted into three color $u$ and three anti-color ${\bar d}$ in high T or QGP phase. So a transition from 1 to 6 quantum states in color space can be realized (extending static to dynamical picture, one can add color gluons also). Similar to color space, when we include flavor/iso-spin, spin spaces, we will get a collective reduction of degeneracy factor from $g=52$ to $g\approx7$ by decreasing the temperature from QGP to hadronic phase.
Reduction of LQCD thermodynamics from their SB limits can be realized by the reduction of degeneracy factor of QGP system. So $g(T)$ can grossly map $T$-dependent QCD interaction. 

Now, the temperature basically measure of randomness of the system. So, low to high T represents less to more quantum states, interpreting less to more randomness. On the other hand, magnetic field plays opposite role as it tries to make the system be more ordered. So
low to high $eB$ represents less to more orderliness or more to less randomness. Owing to that fact, low $T$ and high $eB$ corresponds to hadronic phase with smaller number of degeneracy factors, while, high $T$ and low $eB$ corresponds to QGP phase with larger number of degeneracy factors. Reduction of transition temperature $T_c$ by increasing $eB$ is noticed in LQCD data~\cite{Bali1,Bali2}, which can classify two domains in $T$-$eB$ plane.
In right-middle panel of Fig.~(\ref{gvsT}), reducing of degeneracy factors by increasing $eB$ in low $T$ domain can be noticed. Roughly within $0.1 ~{\rm GeV}<T<0.170 ~{\rm GeV}$ and $0<eB<0.4~ {\rm GeV}^2$, major suppression of degeneracy factors is occurred by decreasing $T$ and increasing $eB$. In this way, $g(T,B)$ can grossly map QCD interaction as a function of $T$ and $eB$.   

\section{Estimation of shear viscosity and electrical conductivity of interacting QGP in presence of magnetic field}
\label{sec:sh_el}
In this section, we will see the role of QCD interaction in presence of magnetic field 
on transport coefficients like shear viscosity and electrical conductivity, where 
the interaction is mapped by quasi-particle description, discussed earlier section. 
The details formalism of shear viscosity and electrical conductivity in presence of
magnetic field are derived in Appendix, given in Sec.~(\ref{sec:App}).

In absence of magnetic field, medium follow isotropic transport properties,
for which we will get single component of shear viscosity ($\eta$) and electrical
conductivity ($\sigma$), but they become multi-component in presence of magnetic field.
We will get five shear viscosity components $\eta_n$ ($n=0,1,..,4$)
and three electrical conductivity components $\sigma_n$ ($n=0,1,2$), which can be
classified into three main components - parallel ($\eta_{\parallel}$, $\sigma_{\parallel}$), perpendicular ($\eta_{\perp}, \sigma_{\perp}$) and Hall ($\eta_{\times}, \sigma_{\times}$) components.
Getting guided from Sec.~(\ref{sec:App}) (Appendix), let us first write the expressions of 
$\eta$ and $\sigma$ for massless QGP system in absence of magnetic field:
\bea
\eta &=&\frac{\beta}{15}g_g\int \frac{d^3\vk}{(2\pi)^3}\frac{\vk^4}{E^2}\tau_cf_0(1+ f_0)
+\frac{g_{Q}\beta}{15}\int \frac{d^3\vk}{(2\pi)^3}\frac{\vk^4}{E^2}\tau_cf_0(1- f_0)
\nn\\
&=&\Big[g_g +\frac{7}{8}g_Q\Big]\frac{4\tau_c\zeta(4)T^4}{5\pi^2}
\label{sh_B0_QGP}
\eea
and 
\bea
\sigma &=&\frac{g_Q}{3}\sum_{f=u,d,s}\frac{{\tilde e}^2_f\beta}{3}\int \frac{d^3\vk}{(2\pi)^3}\frac{\vk^2}{E^2}\tau_cf_0(1- f_0)
\nn\\
&=&\frac{g_Q}{3}\sum_{f=u,d,s}{\tilde e}^2_f\frac{\zeta(2)}{3\pi^2}\tau_c T^2
\label{el_B0_QGP}
\eea
For more simplified case - spinless and massless QGP system (by considering MB distribution function), Above expressions can be written as
\bea
\eta &=&\Big[g_g + g_Q\Big]\frac{4\tau_c T^4}{5\pi^2}
\label{eta_MB_m0}
\\
\sigma &=&\frac{g_Q}{3}\sum_{f=u,d,s}{\tilde e}^2_f\frac{1}{3\pi^2}\tau_c T^2~.
\label{MB_m0}
\eea
Applying quasi-particle description in Eqs.~(\ref{eta_MB_m0}), (\ref{MB_m0}), we can express $\eta$ and $\sigma$ as
\bea
\eta &=&\frac{\tau_c}{15}(g_g+g_Q)\beta\int \frac{d^3\vk}{(2\pi)^3}\Big(\frac{\vk}{E}\Big)^2\vk^2f_0
\nn\\
&\approx&\frac{v^2_{\rm av}k^2_{\rm av}}{15}\tau_c\chi
\label{et_QP_B0}
\\
\sigma &=&\frac{g_Q}{3}\sum_{f=u,d,s}\frac{{\tilde e}^2_f\tau_c}{3}\beta\int \frac{d^3\vk}{(2\pi)^3}\Big(\frac{\vk}{E}\Big)^2f_0
\nn\\
&\approx&\sum_{f=u,d,s}\frac{{\tilde e}^2_f\tau_c}{3}v^2_{\rm av}\chi_e~.
\label{es_QP_B0}
\eea
where different static susceptibilities can be expressed in terms of their massless values as
\bea
\chi &=& g(T)\times\chi(m=0)
\nn\\
\chi_e &=& g(T)\times\chi_e(m=0)
\eea
with
\bea
\chi(m=0) &=& \Big(\frac{\del n}{\del \mu}\Big)_{\mu=0}
=(g_g+g_Q)\beta\int \frac{d^3\vk}{(2\pi)^3}f_0=(g_g+g_Q)\frac{T^2}{\pi^2}
\nn\\
\chi_e(m=0) &=& \frac{g_Q}{3}\beta\int \frac{d^3\vk}{(2\pi)^3}f_0=\frac{g_Q}{3}\frac{T^2}{\pi^2}~.
\eea
We can roughly get back massless Eqs.~(\ref{eta_MB_m0}), (\ref{MB_m0}) from quasi-particle Eqs.~(\ref{et_QP_B0}), (\ref{es_QP_B0}) by putting $v_{\rm av}=1$, $k_{\rm av}=E_{\rm av}=3T$. So we can build a simple analytic expressions
of $\eta$ and $\sigma$ in quasi-particle picture:
\bea
\eta &=& \Big[v_{\rm av}^2(T)f_k^2(T) g(T)\Big](g_g+g_Q)\frac{9T^4}{15\pi^2}\tau_c
\nn\\
\sigma &=& \Big[v_{\rm av}^2(T) g(T)\Big]g_Q\frac{T^2}{9\pi^2}\tau_c\sum_{f=u,d,s}q_f^2~,
\label{eta_sig_Int}
\eea
where the quantities inside the third bracket mainly sources of quasi-particle description, for which estimation will be modified from massless values. This modification basically map the QCD interaction, hidden in LQCD thermodynamics.

Now let us go for finite $B$ picture. From Sec.~(\ref{sec:App}), realizing the anatomy
of parallel and perpendicular components of viscosity and conductivities of charged medium,
\bea
\eta_{\parallel} &=&\frac{\eta}{1+4(\tau_c/\tau_B)^2}
\nn\\
\eta_{\perp} &=&\frac{\eta}{1+(\tau_c/\tau_B)^2}
\nn\\
\sigma_{\parallel} &=&\sigma
\nn\\
\sigma_{\perp} &=&\frac{\sigma}{1+(\tau_c/\tau_B)^2}~,
\eea
we can build quasi-particle expressions of QGP system:
\bea
\eta_{\parallel}(T,B) &=&\Big[v_{\rm av}^2(T,B)f_k^2(T,B) g(T,B)\Big]\frac{9T^4}{15\pi^2}\tau_c\Big[g_g+\frac{g_Q}{3}\sum_{f=u,d,s}\frac{1}{1+4(\tau_c/\tau_{B,f})^2}\Big]
\nn\\
\eta_{\perp}(T,B) &=&\Big[v_{\rm av}^2(T,B)f_k^2(T,B) g(T,B)\Big]\frac{9T^4}{15\pi^2}\tau_c\Big[g_g+\frac{g_Q}{3}\sum_{f=u,d,s}\frac{1}{1+(\tau_c/\tau_{B,f})^2}\Big]
\nn\\
\sigma_{\parallel}(T,B) &=&\Big[v_{\rm av}^2(T) g(T)\Big]g_Q\frac{T^2}{9\pi^2}\tau_c\sum_{f=u,d,s}q_f^2
\nn\\
\sigma_{\perp}(T,B) &=&\Big[v_{\rm av}^2(T) g(T)\Big]g_Q\frac{T^2}{9\pi^2}\tau_c\sum_{f=u,d,s}q_f^2\frac{1}{1+(\tau_c/\tau_{B,f})^2}~,
\eea
where $\tau_{B,f}=\frac{E_{\rm av}}{q_f B}$ is inverse of synchrotron frequency of different flavor $f$ with electric charge $q_f$.

 \begin{figure}
 \centering
 \includegraphics[scale=0.3]{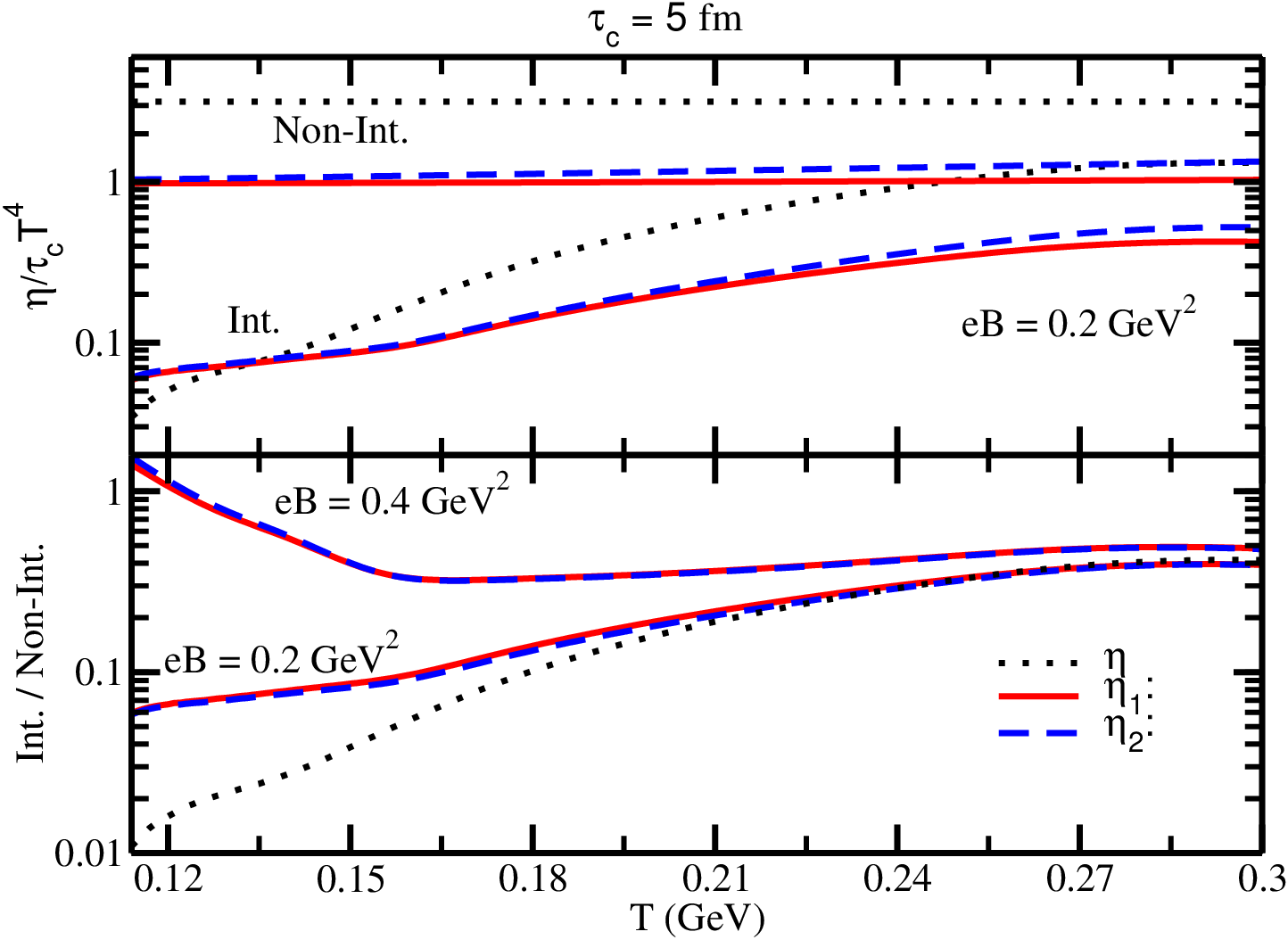}
 \includegraphics[scale=0.3]{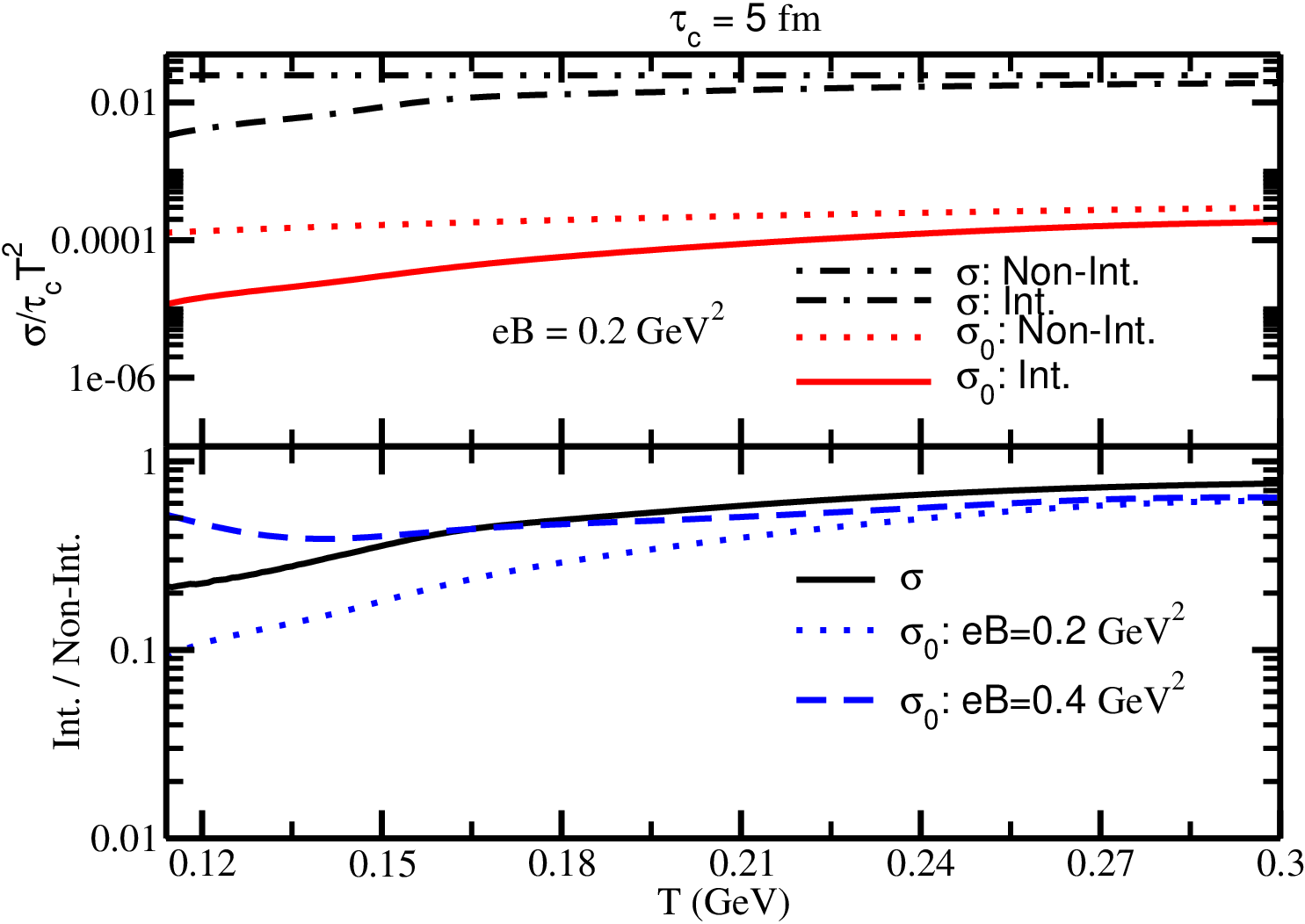} 
 \caption{Left-upper: $T$ dependence of $\eta$ (dotted line), $\eta_1(eB=0.2 {\rm GeV}^2)$ (solid line) and $\eta_2(eB=0.2 {\rm GeV}^2)$ (dash line) for their non-interacting and interacting cases. Left-lower: Ratio of interaction to non-interaction values of $\eta$, $\eta_1$ and $\eta_2$ at different magnetic fields. Right-upper: $T$ dependence of $\sigma$, $\sigma_0(eB=0.2 {\rm GeV}^2)$ for their non-interacting and interacting cases. Right-lower: Ratio of interaction to non-interaction values of $\sigma$, $\sigma_0$ at different magnetic fields.}
 \label{eta_sig_T}
 \end{figure}

Let us come to the numerical estimation.
According to Eqs.~(\ref{et_QP_B0}), (\ref{el_B0_QGP}) or (\ref{eta_MB_m0}), (\ref{MB_m0}), the normalized value $\eta/(\tau_c T^4)$ and $\sigma/(\tau_c T^2)$ for massless QGP will be constant as shown by straight horizontal dotted (Left-upper) and dash double-dotted (Right-upper) lines in Fig.~(\ref{eta_sig_T}).
Next, using Eq.~(\ref{eta_sig_Int}), we will get the values of $\eta/(\tau_c T^4)$ (black dotted line) and $\sigma/(\tau_c T^2)$ (red dotted line) of interacting QGP, which are suppressed from their massless values and also carry an additional $T$-dependent profile due to the quasi-particle $T$-dependent quantities - $v_{\rm av}(T)$, $f_k(T)$ and $g(T)$. The results say that similar to reduced profile of thermodynamical quantities along $T$-axis with respect to their massless or SB limits, transport coefficients will also follow that pattern and the reduction of both quantities increase as one goes high $T$ or perturbative QCD domain to low $T$ or non-perturbative QCD domain. Here, the dissipation information of transport coefficients, hidden in relaxation time $\tau_c$, are normalized to reveal their non-interaction and interaction component of thermodynamical phase space only. Unlike to thermodynamical quantities, transport coefficients basically carry two parts of interactions - one is in thermodynamical phase-space via quasi-particle based tuning parameters and another is in relaxation time, interpreting dissipation component of interaction. A possible connection between two interaction is  attempted in Ref.~\cite{SS_JPG}, where two time scale, coming from thermodynamics and dissipation interactions are compared. Searching the connection between two interactions might be difficult and framework dependent. Hence, without entering to this complicated part, we have take two interactions as two independent components, where former is guided from LQCD data and latter is kept as free by choosing free parameter $\tau_c$. The term ``non-interacting'' might be little misguiding term. Non-interacting dissipation mean $\tau_c\rightarrow \infty$, where transport coefficients are diverged but non-interacting thermodynamical phase-space provide a massless expression. So terms non-interacting and interacting in present article will not applicable for dissipation component, rather only for thermodynamical phase-space component. In both cases (interacting and non-interacting), we will keep relaxation time as finite and free parameter.

By taking ratio of interaction to non-interaction values, as shown in lower panels of Fig.~(\ref{eta_sig_T}), we can see that when we go from $T=0.300$ GeV to $0.120$ GeV, shear viscosity and electrical conductivity can respectively face $50\%$ to $98.5\%$ and $30\%$ to $90\%$ reduction due to interaction.     

Now, at finite $B$ picture, non-interacting results $\eta_{1}$ (red solid line), $\eta_2$ (blue dash line) and $\sigma_0$ (red dotted line) are shown in upper panels of Fig.~(\ref{eta_sig_T}), which shows a reduction with respect to their without field results. One can understand that this reduction comes through the $\tau_B\approx \frac{3T}{q_fB}$, which transform $\tau_c$ to a reduced scale $\frac{\tau_c}{1+\Big(\frac{\tau_c}{\tau_B}\Big)^2}$ roughly. The reduction will increase with increasing of $B$ and decreasing of $T$.
Going to interaction picture at finite $B$, we will get further modification and we have to use now $T$ and $B$-dependent quantities - $v_{\rm av}(T,B)$, $f_k(T,B)$ and $g(T,B)$. The results are shown by red solid, blue dash lines for $\eta_1$, $\eta_2$ in left-upper panel of Fig.~(\ref{eta_sig_T}) and by red solid line for $\sigma_o$ in right-upper panel of Fig.~(\ref{eta_sig_T}). So finite magnetic field will first change the values of transport coefficients via $\tau_B(T,B)$ and then interaction will change their values further through $v_{\rm av}(T,B)$, $f_k(T,B)$ and $g(T,B)$. The $T$, $B$ dependence of interaction can be visualized better way in lower panels of Fig.~(\ref{eta_sig_T}), which reveal a less reduction in low $T$ and high $B$ domain due to interaction as observed in thermodynamical quantities also. Another point - anisotropy in dissipation can be realized through the inequality parallel $>$ perpendicular component - $\eta_1 > \eta_2$ and $\sigma > \sigma_0$.

  \begin{figure}
 \centering
 \includegraphics[scale=0.3]{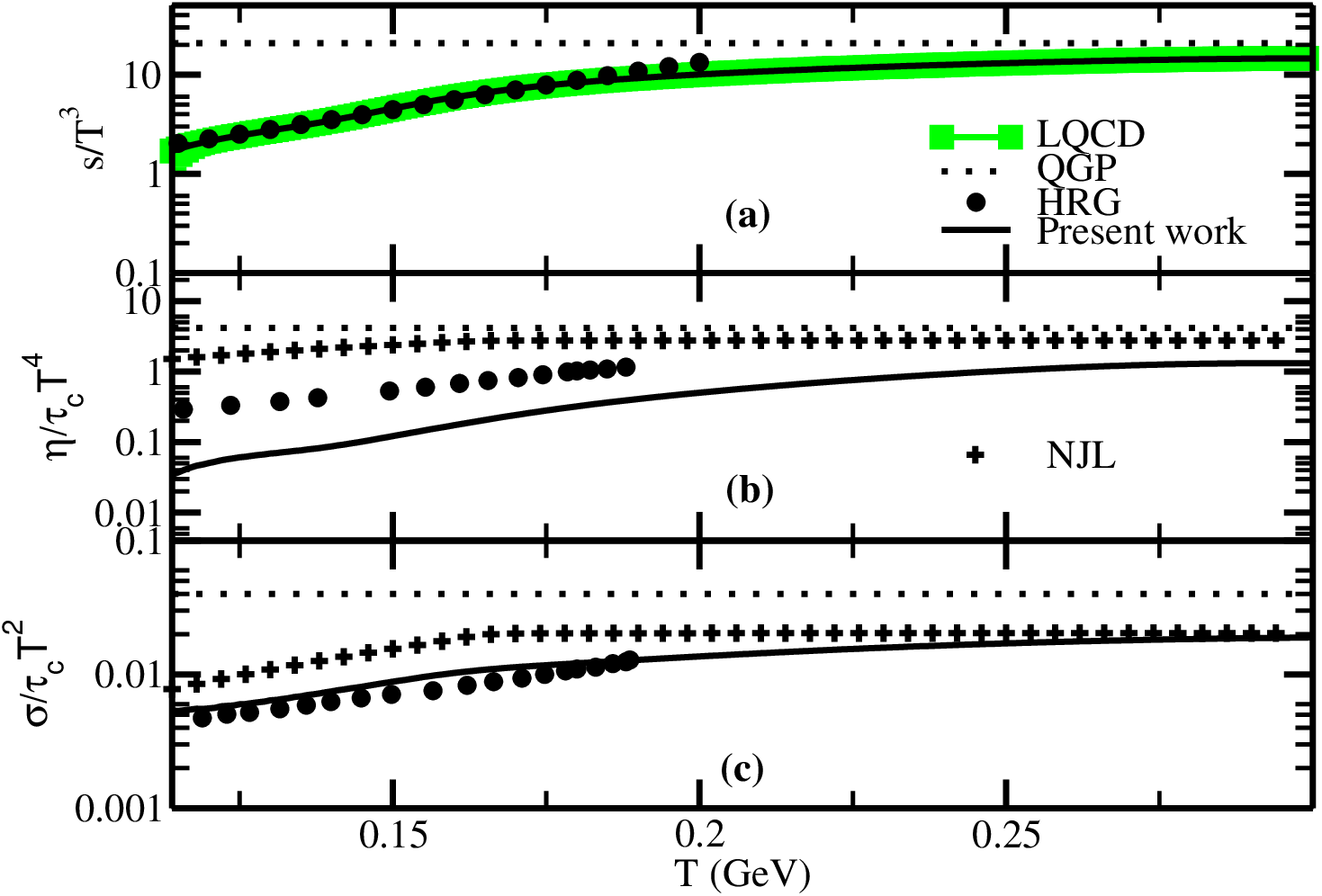}
 \includegraphics[scale=0.3]{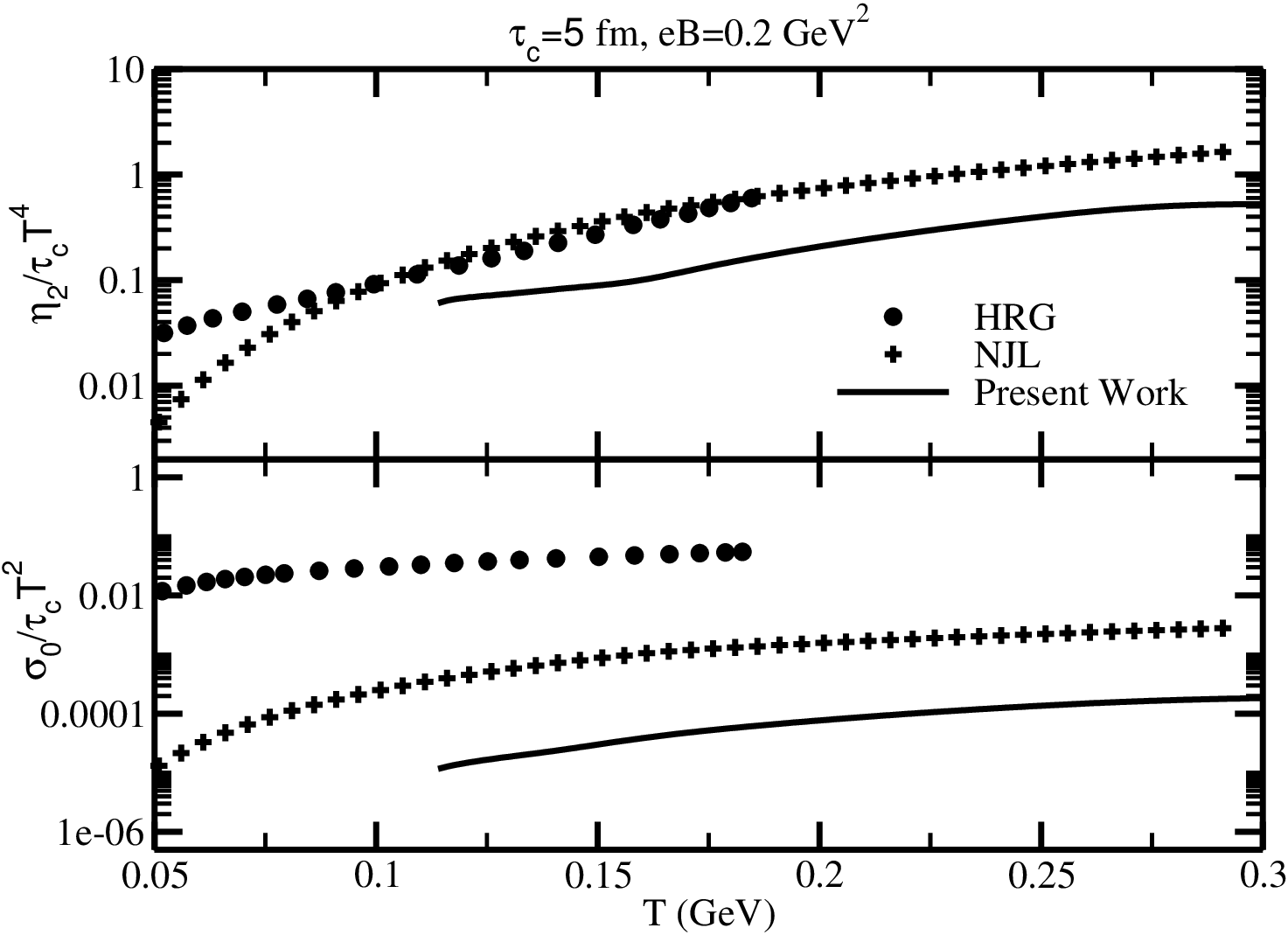} 
 \caption{Left: Comparing our estimation (black solid line) of entropy density (upper panel), shear viscosity (middle panel) and electrical conductivity (lower panel) with the corresponding values of (ideal) HRG model~\cite{HRGB}, NJL model~\cite{NJLB_el,NJLB_eta} and massless limits (dotted line). Green squares are LQCD data~\cite{Bali2} of entropy density. Right: perpendicular components of shear viscosity (upper panel) and electrical conductivity (lower panel) in our model, HRG model~\cite{HRGB}, NJL model~\cite{NJLB_el,NJLB_eta} at $eB=0.2$ GeV$^2$ and $\tau_c=5$ fm.}
 \label{match_TB}
 \end{figure}
\subsection{Comparing interaction strength with earlier estimations}
In this section, we will try to explore how close/far our estimation of transport coefficients with other model-dependent estimations, which can map the interaction of LQCD thermodynamics. In Fig.~(\ref{match_TB}), left-upper panel shows that the LQCD data~\cite{Bali2} (green squares), our fitted curve (black solid line) and (ideal) HRG model estimation~\cite{HRGB} (black circles) for $s/T^3$ at $eB=0$ are well agreement. Extracting the interaction information from LQCD thermodynamics via quasi-particle $T$-dependent quantities $v_{\rm av}(T)$, $f_k(T)$ and $g(T)$, we have basically projecting them to transport coefficients. We have compared our estimations
of $\eta/(\tau_cT^4)$ and $\sigma/(\tau_cT^2)$ with HRG~\cite{HRGB} (circles) and NJL~\cite{NJLB_el,NJLB_eta} (pluses) model estimations in middle and lower panels of Fig.~(\ref{match_TB}).
All results are suppressed from their SB or massless limits, shown by dotted horizontal line.
Suppression values of $s/T^3$, $\eta/(\tau_cT^4)$ and $\sigma/(\tau_cT^2)$ are mapping their non-pQCD estimations. Qualitative trends of different models are quite similar with little quantitative differences in transport coefficients.
At finite $B$, these quantitative differences are revealing more as shown in the right panel of Fig.~(\ref{gvsT}). From Ref.~\cite{Bali2}, we notice that finite $B$ extension of $s$ in HRG model is not well matched with corresponding LQCD values, which is also true from NJL model estimations of Refs.~\cite{PB4_Farias,NJLB_el}. In that context our estimations is purely projected from LQCD thermodynamics in presence of magnetic field. So $T$, $B$ dependence of our estimated transport coefficients might carry more accurate $T$, $B$ dependent non-perturbative QCD interaction. 
\begin{figure}
 \centering
 \includegraphics[scale=0.35]{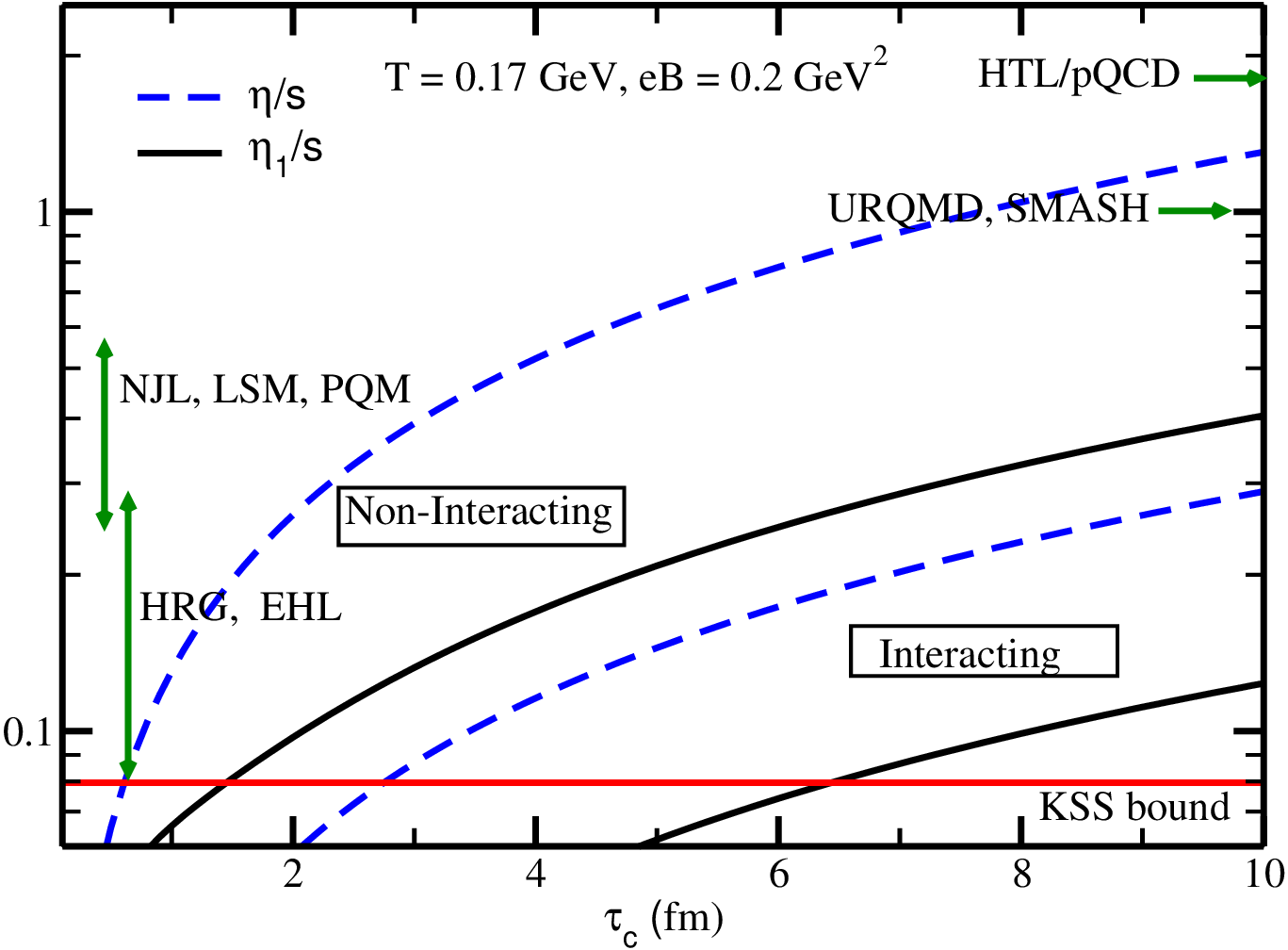}
 \caption{Viscosities to entropy density ratio for non-interacting and interacting QGP at eB=0 (dash line), $eB=0.2$ GeV$^2$ (solid line). Green arrows for earlier estimations at $eB=0$.}
 \label{etabys}
 \end{figure}

Let us now focus on famous dimensionless quantity - ratio between shear viscosity
and entropy density, which measure the fluid nature of the medium. From experimental
side, this quantity should be very close to KSS bound $1/(4\pi)$~\cite{KSS}, which is
drawn by red solid (horizontal) line in Fig.~(\ref{etabys}). At $B=0$ case, a long list of references can be found on the estimations of $\eta/s$, where few~\cite{Arnold,Marty1,Sasaki2,G_IFT,Kadam_NJL,Purnendu,HM_PQM,Bass,Juan,SG_piN_IFT,SS_piNK,SG_HRG,AD_HRG,GPK_HRG} are tabulated in Table~[1].  
\begin{table} 
\caption{Order of magnitude of $\eta/s$ from different model calculations (first column) with 
references at temperature range below (second column)
and above (third column) transition temperature $T_c$.}
\begin{center}
\begin{tabular}{l|cc}
Framework [Reference] & $T ~\leq~T_c$ & $T ~\geq~T_c$  \\
\hline
HTL~\cite{Arnold} & - & $1.8$  \\
NJL~\cite{Marty1} & $1$-$0.3$ & $0.3$-$0.08$  \\
NJL~\cite{Sasaki2} & $1$-$0.5$ & $0.5$-$0.55$  \\
NJL~\cite{G_IFT} & - & $0.5$-$0.12$  \\
NJL~\cite{Kadam_NJL} & $2$-$0.25$ & $0.25$-$0.5$  \\
LSM~\cite{Purnendu} & $0.87$-$0.55$ & $0.55$-$0.62$  \\
PQM~\cite{HM_PQM} & $5$-$0.5$ & $0.3$-$0.08$  \\
URQMD~\cite{Bass} & $1$ & -  \\
SMASH~\cite{Juan} & $1$ & -  \\
EHL~\cite{SG_piN_IFT} & $0.4$-$0.1$ & -  \\
EHL~\cite{SS_piNK} & $0.8$-$0.25$ & -  \\
HRG~\cite{SG_HRG} & $0.13$-$0.28$ & -  \\
K-matrix HRG~\cite{AD_HRG} & $0.3$-$0.08$ & -  \\
K-matrix HRG~\cite{GPK_HRG} & $0.4$-$0.08$ & -  
\\
\hline
\end{tabular}
\end{center}
\label{tab1}
\end{table}
Relaxation time $\tau_c$ are calculated in different models and they get order of magnitude of $\eta/s$ but in present article, we keep it as parameter. By plotting $\eta/s$ against $\tau_c$-axis for non-interacting and interacting case, we can get an effective $\tau_c$ ranges, within which earlier estimated values are located. One can find that pQCD~\cite{Arnold} and box-simulation~\cite{Bass,Juan} based on URQMD~\cite{Bass}, SMASH~\cite{Juan} codes are showing 10-20 larger values than KSS bound. Whereas effective QCD models~\cite{Marty1,Sasaki2,G_IFT,Kadam_NJL,Purnendu,HM_PQM} like NJL~\cite{Marty1,Sasaki2,G_IFT,Kadam_NJL}, LSM~\cite{Purnendu}, QM~\cite{HM_PQM} models addressed an intermediate ranges $\eta/s\approx 2.5-5.5$ (marked by green arrow). Effective hadron Lagrangian (EHL) models as well as HRG models provide the range $\eta/s\approx 2.5-0.08$ (marked by green arrow), where K-matrix HRG model provide KSS values $0.08$ near transition temperature $T\approx 0.170$ GeV.
After realizing the range of $\eta/s$ within 1-20 times KSS bound can be found for $\tau_c\approx 0.5-10$ fm, we have shown their finite $B$ extended results at $eB=0.2$ GeV$^2$. Due to log scale of $\eta/s$-axis, it is not clear but in linear scale one can see $\eta/s\propto \tau_c$ for non-interacting and $eB=0$ case and its crossing point with KSS line has $T$-dependence - $\tau_c(T)=5/(4\pi T)$. For interacting picture, still linear relation $\eta/s\propto \tau_c$ maintained but it get suppressed due to $T$-dependent interaction information, entered through $g(T)$, $E_{\rm av}(T)$ etc. In presence of magnetic, quark component follow $\frac{\eta_1}{s}\propto\frac{\tau_c}{1+4(\tau_c/\tau_B)^2}$, for which a non-linear trend can be expected but gluon component remain same as without field case.
So, when we add gluon and quark components, a mild non-linear nature is observed. 

\subsection{Merits, Demerits and Outlook}
Present work has attempted to map LQCD thermodynamics in presence of magnetic field~\cite{Bali1,Bali2} by using $T$, $B$ dependent tuning parameters - degeneracy factor $g(T,B)$, average energy $E_{\rm av}(T,B)$, momentum $k_{\rm av}(T,B)$, velocity $v_{\rm av}(T,B)$ of the constituents of QGP system. Now for massless (SB limit) and $B=0$ case, their values are $g(T, B=0)=1$ (which is basically the maximum limit of the fraction, multiplied with total degeneracy factor 52 of QGP system), $E_{\rm av}(T, B=0)=k_{\rm av}(T, B=0)=3T$ and $v_{\rm av}(T, B=0)=1$. In presence of QCD interaction, $g(T, B=0)<1$, $E_{\rm av}(T, B=0)<1$, $k_{\rm av}(T, B=0)<3T$ and $v_{\rm av}(T, B=0)<1$, which are observed during tuning these parameters to fit the LQCD data. When external magnetic field is applied to the interacting QCD system, the $g(T, B)$ face further suppression (in hadronic temperature range) and $E_{\rm av}(T, B)$, $k_{\rm av}(T, B)$ and $v_{\rm av}(T, B)$ face enhancement with a non-monotonic $T$-profile. So this picture is quite simple and no quantum effect via Landau quantization has been considered, which might be the demerit part of the present model. Whereas simple picture with analytic type expression might be considered as the merit part of the model. Though the 
methodology is not so robust, but with the help of its simple parametric expressions, one can get a quick rough estimation of any phenomenological
quantity, influenced by temperature and magnetic field dependent QCD interaction. 
For example, Feynman diagrams connected with QGP signal like strange enhancement, thermal dilepton and photon production, heavy quark diffusion, jet quenching parameters etc. might be calculated in quasi-particle kinematics and phase-space, which might provide us a non-perturbative estimations. However, actual quantitative calculations of different signal related phenomenology needs more details knowledge, which may be presented elsewhere.

\section{Summary}
\label{sec:sum}
The present article is first aimed to map the QCD interaction in a very simplified way. 
It is LQCD simulation, which has provided the QCD interaction at finite 
temperature and magnetic field by calculating thermodynamical quantities
of QGP, which always remain lower than their non-interacting or massless values, popularly called SB limits. Along with the thermodynamics, LQCD calculation
has found temperature and magnetic field dependent of quark condensate, which exposes inverse magnetic catalysis phenomenon near quark-hadron transition temperature. Constructing constitutent quark mass, proportional to the temperature and magnetic field dependent quark condensate of LQCD, we have first obtained total number density of QGP, which is certainly deviated from its massless or SB limit. Assuming degeneracy factor as one of the tuning parameter, we have matched LQCD based total number density, which indirectly map the quark condensate of LQCD. Next, to match the LQCD   
thermodynamical quantities like pressure, energy density, entropy density, we have considered a temperature and magnetic field dependent average energy, momentum and velocity of partons (quarks/gluons) and tuned them accordingly.

After mapping temperature and magnetic field dependent of QCD interaction,
we have applied it to calculate 
transport coefficients like shear viscosity and electrical conductivity
in presence of magnetic field. Isotropic property of medium is generally broken due to
magnetic field and hence, single component transport coefficient
are splitted into multi-components, which can be classified into two main
components - parallel and perpendicular components of transport coefficients. There can be Hall component at non-zero quark chemical potential but since here our focus on zero chemical potential zone, so net Hall component will be disappeared due to cancellation of quark and anti-quark contribution. 
Our final expressions of transport coefficients carry two time scales, originated from collision and magnetic field, where latter one Completely vanished at zero magnetic field case. So the differences among different component of transport coefficients are coming due to their different functional dependence with the magnetic time scale, which is basically inverse of synchrotron frequency of quarks.
Qualitatively, we find a overall suppression of transport coefficient because of magnetic
field and then when we plug in the LQCD interaction, they get further suppression.
Based on the present study, one can identify the magnetic field and interaction as
two dominating sources, for which the transport coefficients of QGP become lower. The well famous - low values shear viscosity to entropy density ratio might be linked with these two sources. This understanding might be better when a magneto-hydrodynamic simulation provide the values of shear viscosity to entropy density ratio by matching relevant experimental data.


%
{\bf Acknowledgment:} SG and JD acknowledge to
MHRD funding via IIT Bhilai and SS acknowledges to
facilities, provided from IIT Bhilai in self-sponsor PhD scheme. 
AM and SP thank for (payment basis) hospitality from 
IIT Bhilai during his summer internship tenure (May-June, 2019).

\section{Appendix}
\label{sec:App}
\subsection{Shear viscosity calculation in presence of magnetic field}
\label{sec:Sh_B}
Let us consider a relativistic fermion/boson fluid, whose dissipative
energy momentum tensor $\Delta T_{\mu\nu}$ due to shear stress is connected
with velocity gradient-type tensor
\bea
U_{\mu\nu} &=& D^\mu u^\nu + D^\nu u^\mu +\frac{2}{3}
\Delta^{\mu\nu}\partial_\sigma u^\sigma~~~\mbox{with}
\nn\\
D^\mu &=& \partial^\mu - u^\mu u^\sigma \partial_\sigma,~
\Delta^{\mu\nu}=u^\mu u^\nu - g^{\mu\nu}~.
\eea
via macroscopic relation
\be
\Delta T_{\mu\nu} = \eta_{\mu\nu\alpha\beta} U^{\alpha\beta}~,
\label{macro_Sh}
\ee
where $\eta_{\mu\nu\alpha\beta}$ is shear viscosity tensor, which is aimed to
estimated microscopically in this section. Assuming fermion/boson
equilibrium distribution function
\be
f_0=\frac{1}{e^{\beta\om}\mp 1}
\ee
get deviation
\bea
\delta f &=& -\phi\Big(\frac{\del f_0}{\del \om}\Big)
\nn\\
&=& -A(k_\mu k_\nu U^{\mu\nu})\Big(\frac{\del f_0}{\del \om}\Big)
\nn\\
&=& A(k_\mu k_\nu U^{\mu\nu})\beta f_0(1\mp f_0)~,
\label{d_fsh}
\eea
we can get microscopic expression of (dissipative) energy-momentum
\bea
\Delta T_{\mu\nu} &=&g\int \frac{d^3\vk}{(2\pi)^3}\frac{k_\mu k_\nu}{\om}\delta f
\nn\\
&=&g\beta\int \frac{d^3\vk}{(2\pi)^3}\frac{k_\mu k_\nu }{\om}k_\alpha k_\beta U^{\alpha\beta}A f_0(1\mp f_0)~,
\label{micro_sh}
\eea
where $\om=\{\vk^2+m^2\}^{1/2}$ is energy and $g$ is
degeneracy factor of fermion/boson. 
To determine unknown constant $A$, we use
the relaxation time approximation (RTA) in Boltzmann equation,
\bea
\frac{\partial f}{\partial t} +\frac{\del x^j}{\del t}\frac{\partial f}{\partial x^j} 
+ \frac{\del k_j}{\del t}\frac{\partial f}{\partial k^j} &=& {\cal C}[\delta f]
\nn\\
\frac{\partial f}{\partial t} +\frac{k^j}{\om}\frac{\partial f}{\partial x^j} &=& \frac{\delta f}{\tau_c}
\nn\\
\Rightarrow \delta f &=&\frac{\tau_c}{\om}k^\mu\del_\mu f_0 ({\rm since},~ \frac{\del k_j}{\del t}=0)
\nn\\
&=& \frac{\tau_c}{\om}k^\mu k^\nu U_{\mu\nu} \beta f_0(1\mp f_0)
\label{Boltz_df}
\eea
Comparing Eq.~(\ref{Boltz_df}) and (\ref{d_fsh}), one can identify the unknown constant
\be
A=\frac{\tau_c}{\om}
\ee
After knowing $A$, the full connection between macroscopic Eq.~(\ref{macro_Sh}) 
and microscopic Eq.~(\ref{micro_sh}) can be written as
\bea
\eta_{\mu\nu\alpha\beta}U^{\alpha\beta}&=&\Delta T^{\mu\nu}
=g\int \frac{d^3\vk}{(2\pi)^3}\frac{k_\mu k_\nu}{\om}\delta f
\nn\\
&=&\Big\{g\beta\int \frac{d^3\vk}{(2\pi)^3}\frac{k_\mu k_\nu k_\alpha k_\beta}{\om^2}\tau_cf_0(1\mp f_0)\Big\}U^{\alpha\beta}
\nn\\
\Rightarrow \eta &=& \frac{g\beta}{15}\int \frac{d^3\vk}{(2\pi)^3}\frac{\vk^4}{\om^2}\tau_cf_0(1\mp f_0)~,
\label{eta_B0}
\eea
where we have used the angular average (denoted by $\langle ..\rangle_\theta$) identity
\be
\langle k_{\alpha}k_{\beta}k_{\gamma}k_{\delta} \rangle_\theta  
= \frac{k^4}{15}(\delta_{\alpha\beta}\delta_{\gamma\delta} 
+ \delta_{\alpha\gamma}\delta_{\beta\delta}+ \delta_{\alpha\delta}\delta_{\beta\gamma})~.
\ee

The isotropic property of shear viscosity $\eta$ does not hold in presence of magnetic
field. Instead of single $\eta$, we will get many component $\eta_n$, which mean that
different directional shear stress become different in presence of magnetic field.
Following the prescriptions of Ref.~\cite{Landau,JD_eta},
the dissipative part of the energy-momentum tensor in presence of magnetic field $B$ can be expressed as
\bea
\Delta T_{\alpha\beta} = \sum_{n=0}^{4}\eta_nV_{\alpha\beta}^n
\eea
where $\eta_n$ with $n=0, 1, 2, 3, 4$ are five components of viscosities
and the velocity gradient-type tensor $V^n_{\mu\nu}$ are taken same as addressed in Ref.~\cite{Landau,Tuchin}:
\bea
&&V_{\alpha\beta}^0 = (3b_{\alpha}b_{\beta}-\delta_{\alpha\beta})(b_{\gamma}b_{\delta}V_{\gamma\delta}-\frac{\vec{\nabla}.\vec{v}}{3})\\
&&V_{\alpha\beta}^1 = 2V_{\alpha\beta} + \delta_{\alpha\beta}V_{\gamma}b_{\gamma}b_{\delta}-2V_{\alpha\gamma} b_{\gamma}b_{\beta}-2V_{\beta\gamma}b_{\gamma}
b_{\alpha}(b_{\alpha}b_{\beta}-\delta_{\alpha\beta})\vec{\nabla}.\vec{v}\nn \\
&& + b_{\alpha}b_{\beta}V_{\alpha\delta}b_{\gamma}b_{\delta}\\
&& V_{\alpha\beta}^2 = 2(V_{\alpha\gamma}b_{\beta}b_{\gamma} + V_{\beta\gamma}b_{\alpha}b_{\gamma}-2b_{\alpha}b_{\beta}V_{\gamma\delta}b_{\gamma}b_{\delta})\\
&& V_{\alpha\beta}^3 = V_{\alpha\gamma}b_{\beta\gamma} + V_{\beta\gamma}b_{\alpha\gamma}- V_{\gamma\delta}b_{\alpha\gamma}b_{\beta}b_{\delta}-V_{\gamma\delta}
b_{\beta\gamma}b_{\alpha}b_{\delta}\\
&& V_{\alpha\beta}^4 = 2(V_{\gamma\delta}b_{\alpha\gamma}b_{\beta}b_{\delta}+ V_{\gamma\delta}b_{\beta\gamma}b_{\alpha}b_{\delta})
\eea
with $b_\alpha=B_\alpha/B$, $b_{\alpha\beta}=\ep_{\alpha\beta\mu}B^\mu/B$,
$V_{\alpha\beta} =\frac{1}{2}\Big[\frac{\del u_{\alpha}}
{\del x_{\beta}} + \frac{\del u_{\beta}}{\del x_{\alpha}}\Big]$~.
Let us assume our deviation $\delta f$ from equilibrium accordingly:
\bea
\delta f &=& -\phi\Big(\frac{\del f_0}{\del \om}\Big)
\nn\\
&=& -\sum_{n=0}^4 A_n(k_\mu k_\nu U^{\mu\nu}_n)\Big(\frac{\del f_0}{\del \om}\Big)
\nn\\
&=& \sum_{n=0}^4 A_n(k_\mu k_\nu U^{\mu\nu}_n)\beta f_0(1\mp f_0)~,
\label{d_fshB}
\eea
To determine unknown constant $A$, we use
the relaxation time approximation (RTA) in relativistic Boltzmann equation (RBE) at finite $B$,
\bea
\frac{\partial f}{\partial t} +\frac{\del x^j}{\del t}\frac{\partial f}{\partial x^j} 
+ \frac{\del k_j}{\del t}\frac{\partial f}{\partial k^j} &=& {\cal C}[\delta f]
\nn\\
\frac{1}{\om} k^\mu\del_\mu f_0 + \frac{{\tilde e}B}{\om}b_{\alpha\beta}k_{\beta}\frac{\del (\delta f)}{\del k_{\alpha}}
&=& \frac{\delta f}{\tau_c}~,
\label{Boltz_dfB}
\eea
where force term in RBE can not contribute with $f_0$, therefore, we have to proceed for $\delta f$
order contribution. In other way, magnetic relaxation time $\tau_B=\om/({\tilde e}B)$ along
with collisional relaxation time $\tau_c$ are responsible for deviation $\delta f$.
Now, by using Eq.~(\ref{d_fshB}) in (\ref{Boltz_dfB}), we get 
\bea
&&\frac{1}{\om}k_{\alpha}k_{\beta}V_{\alpha\beta}\{\beta f_0(1\mp f_0)\} 
+\frac{b_{\alpha\beta}k_{\beta}}{\tau_B}\Big(\sum_{n=0}^4 A_n V_{\alpha\gamma}^n k_{\gamma}\Big)
\{\beta f_0(1\mp f_0)\}
\nn\\
&&=
-\frac{1}{\tau_c}\sum_{n=0}^4 A_n V_{\gamma\delta}^n k_{\gamma}k_{\delta}\{\beta f_0(1\mp f_0)\}
\label{Boltz_An}
\eea

Using some identities~\cite{Landau,Tuchin,JD_eta} and then compare the tensor structures on both sides
to obtain $A_n$ as follows
\bea
&&A_1 = \frac{\tau_c}{\om}\frac{1}{4(1+\frac{\tau_c^2}{\tau_B^2})}\\
&&A_2 = \frac{\tau_c}{\om}\frac{1}{(1+\frac{\tau_c^2}{\tau_B^2})}\\
&&A_3 = \frac{\tau_c}{\om}\frac{(\frac{\tau_c}{\tau_B})}{2(1/4+\frac{\tau_c^2}{\tau_B^2})}\\
&&A_4 = \frac{\tau_c}{\om}\frac{(\frac{\tau_c}{\tau_B})}{(1+\frac{\tau_c^2}{\tau_B^2})}\\
\eea
Putting the $A_n$'s in the expression for $\eta_n$ we get
\bea
&&\eta_1 = \frac{g\beta}{15}\int\frac{d^3\vk}{(2\pi)^3}\frac{\vk^4}{\omega^2}\frac{\tau_c}{4(\frac{1}{4}+\frac{\tau_c^2}{\tau_B^2})}f_0(1\mp f_0)\\
&&\eta_2 = \frac{g\beta}{15}\int\frac{d^3\vk}{(2\pi)^3}\frac{\vk^4}{\omega^2}\frac{\tau_c}{(1+\frac{\tau_c^2}{\tau_B^2})}f_0(1\mp f_0)\\
&&\eta_3 = \frac{g\beta}{15}\int\frac{d^3\vk}{(2\pi)^3}\frac{\vk^4}{\omega^2}\frac{\tau_c(\frac{\tau_c}{\tau_B})}{2(1/4+\frac{\tau_c^2}{\tau_B^2})}f_0(1\mp f_0)\\
&&\eta_4 = \frac{g\beta}{15}\int\frac{d^3\vk}{(2\pi)^3}\frac{\vk^4}{\omega^2}\frac{\tau_c(\frac{\tau_c}{\tau_B})}{(1+\frac{\tau_c^2}{\tau_B^2})}f_0(1\mp f_0)\\
\eea
One can identify parallel, perpendicular and Hall components as $\eta_{xzxz}=\eta_{\parallel}=\eta_2$, $\eta_{xyxy}=\eta_{\perp}=\eta_1$ and $\eta_{\times}=\eta_4$. If we take $B\rightarrow 0$ limit
then $\eta_{\parallel}$ and $\eta_{\perp}$ will be merged to its isotropic value $\eta$ and $\eta_{\times}$ will be disappeared.

\subsection{Electrical conductivity calculation in presence of magnetic field}
\label{sec:el_B}
Let us consider a relativistic fermion/boson fluid, carrying dissipative
current density $J_i$ due to electric field $E^j$ and they are connected
via macroscopic Ohm's law
\be
J_i=\sigma_{ij}E^j~,
\label{macro_E}
\ee
where $\sigma_{ij}$ is conductivity tensor, which is aimed to
estimated microscopically in this section. Assuming fermion/boson
equilibrium distribution function
\be
f_0=\frac{1}{e^{\beta\om}\mp 1}
\ee
get deviation
\bea
\delta f &=& -\phi\Big(\frac{\del f_0}{\del \om}\Big)
\nn\\
&=& -\alpha(k_jE^j)\Big(\frac{\del f_0}{\del \om}\Big)
\nn\\
&=& \alpha(k_jE^j)\beta f_0(1\mp f_0)~,
\label{d_f}
\eea
we can get microscopic expression of (dissipative) current density 
\bea
J_i &=&g{\tilde e}\int \frac{d^3\vk}{(2\pi)^3}\frac{k_i}{\om}\delta f
\nn\\
&=&g{\tilde e}\beta\int \frac{d^3\vk}{(2\pi)^3}\frac{k_ik_j}{\om}\alpha f_0(1\mp f_0)~,
\label{micro_E}
\eea
where ${\tilde e}$ is electric charge, $\om=\{\vk^2+m^2\}^{1/2}$ is energy and $g$ is
degeneracy factor (excluding charge-flavor degeneracy) of fermion/boson. 
To determine unknown constant $\alpha$, we use
the Boltzmann equation,
\be
\frac{\partial f}{\partial t} +\frac{\del x^j}{\del t}\frac{\partial f}{\partial x^j} 
+ \frac{dk_j}{dt}\frac{\partial f}{\partial k^j} = {\cal C}[\delta f]~.
\label{Boltz_B0}
\ee
In the electric-charge-transport picture, the external electric field is responsible 
to make the system deviate from equilibrium. Hence electric force $-{\tilde e}E_j=\frac{dk_j}{dt}$
will build the deviation $\delta f$ and in relaxation time approximation (RTA), we may
assume ${\cal C}[\delta f]=-\delta f/\tau_c$, where  $\tau_c$ is the relaxation time,
required for the system to approach from non-equilibrium to equilibrium state. 
So, Eq.~(\ref{Boltz_B0}) becomes
\bea
-{\tilde e}E_j\frac{\del f_0}{\del k^j}&=&-\delta f/\tau_c
\nn\\
\Rightarrow \delta f &=& \tau_c{\tilde e}E_j\Big(\frac{\del \om}{\del k^j}\Big)
\Big[\frac{\del f_0}{\del \om}\Big]
\nn\\
&=&\tau_c{\tilde e}E_j\Big(\frac{k^j}{\om} \Big)[\beta f_0(1\mp f_0)]
\label{del_f}
\eea
comparing Eq.~(\ref{del_f}) and (\ref{d_f}), one can identify the unknown constant
\be
\alpha=\frac{{\tilde e}\tau_c}{\om}
\ee
After knowing $\alpha$, the full connection between macroscopic Eq.~(\ref{macro_E}) 
and microscopic Eq.~(\ref{micro_E}) can be written as
\bea
\sigma^{ij}E_j&=&J^i_D=g{\tilde e}\int \frac{d^3\vk}{(2\pi)^3}\frac{k^i}{\om}\delta f
\nn\\
&=&\Big\{g{\tilde e}^2\beta\int \frac{d^3\vk}{(2\pi)^3}\frac{k^ik^j}{\om^2}\tau_cf_0(1\mp f_0)\Big\}E_j
\nn\\
\Rightarrow \sigma^{ij}&=&g{\tilde e}^2 \beta\int \frac{d^3\vk}{(2\pi)^3}\tau_c\frac{k^ik^j}{\om^2} f_0(1\mp f_0)
\nn\\
\Rightarrow \sigma &=&\frac{1}{3} g{\tilde e}^2 \beta\int \frac{d^3\vk}{(2\pi)^3}
\tau_c\frac{\vk^2}{\om^2} f_0(1\mp f_0)~.
\label{sigma_B0}
\eea

Next, we will proceed to derive the electrical conductivity in presence of magnetic field $B$,
which is well addressed in Ref.~\cite{Sedrakian_el}.
Here, force term becomes 
$\frac{d\vk}{dt}=-{\tilde e} ({\vec E} +{\vec v}\times {\vec B})$ and so, 
the Boltzmann equation (\ref{Boltz_B0}) becomes
\bea
{-\tilde e} ({\vec E} +\frac{\vec k}{\om}\times {\vec B})\cdot\nabla_k f_0 &=& {\cal C}[\delta f]
\nn\\
{-\tilde e} ({\vec E} +\frac{\vec k}{\om}\times {\vec B})\cdot\Big(\frac{\vk}{\om}\Big)\frac{\del f_0}{\del\om} 
&=& \frac{-\delta f}{\tau_c}~.
\eea
Since the second term of left had side is only magnetic field dependent term and it will be vanished
(following vector identity $(\vk\times{\vec B})\cdot \vk={\vec B}\cdotp(\vk\times\vk)=0$), so we consider the $\nabla_k(\delta f)$ term also 
\be
{-\tilde e}{\vec E}\cdot\Big(\frac{\vec k}{\om}\Big)\frac{\partial f_0}{\partial \om} 
- {\tilde e}(\frac{\vec k}{\om}\times {\vec B})\cdot \nabla_k(\delta f)
 =-\delta f/\tau_c~,
\label{RBE_H_df}
\ee
where we assume $\delta f=-\phi\frac{\del f_0}{\del \om}$ with $\phi=\vk\cdot {\vec F}$ and 
${\vec F}= (\alpha {\hat e} + \beta{\hat h} + \gamma({\hat e}\times{\hat h}) )$, ${\hat e}$ and 
${\hat h}$ are unit vector along electric and magnetic field directions.
Since
\bea
(\frac{\vec k}{\om}\times {\vec B})\cdot\nabla_k(\delta f)
&=&-(\frac{\vec k}{\om}\times {\vec B})\cdot \nabla_k(\vk\cdot {\vec F}) \frac{\partial f_0}{\partial \om}
\nn\\
&=&-(\frac{\vec k}{\om}\times {\vec B})\cdot{\vec F}) \frac{\partial f_0}{\partial \om}
\nn\\
&=&-\frac{\vec k}{\om}\cdot ({\vec B}\times{\vec F}) \frac{\partial f_0}{\partial \om}~,
\label{del_f_expand}
\eea
so Eq.~(\ref{RBE_H_df}) becomes
\bea
\Big(\frac{\vec k}{\om}\Big)\cdot\Big[-{\tilde e}{\vec E} + {\tilde e}({\vec B}\times{\vec F})\Big]
&=&\vk\cdot {\vec F}/\tau_c~,
\nn\\
\frac{1}{\om}\Big[-{\tilde e}E{\hat e} + {\tilde e}B{\hat h}\times(\alpha {\hat e} + \beta{\hat h} + \gamma({\hat e}\times{\hat h}))\Big]
&=&(\alpha {\hat e} + \beta{\hat h} + \gamma({\hat e}\times{\hat h}))/\tau_c
\nn\\
\Big(-\frac{\tau_c{\tilde e}E}{\om} \Big){\hat e} - \Big(\frac{\tau_c{\tilde e}B\alpha}{\om} \Big)({\hat e}\times{\hat h})
+\Big(\frac{\tau_c{\tilde e}B\gamma}{\om} \Big)\{{\hat e}-({\hat e}\cdot{\hat h}){\hat h}\}
&=&(\alpha {\hat e} + \beta{\hat h} + \gamma({\hat e}\times{\hat h}))
\label{Bolz_phi_B}
\eea
%
%
Equating the coefficients of ${\hat e}$, ${\hat h}$ and $({\hat e}\times{\hat h})$ of Eq.~(\ref{Bolz_phi_B}), 
we get
\bea
\Big(\frac{-\tau_c{\tilde e}E}{\om}+\frac{\tau_c\gamma}{\tau_B} \Big)&=&\alpha
\nn\\
\Big(\frac{\tau_c\gamma}{\tau_B} \Big)({\hat e}\cdot{\hat h})&=&\beta
\nn\\
- \Big(\frac{\tau_c\alpha}{\tau_B} \Big)&=&\gamma~,
\eea
where $\tau_B=\om/(eB)$ is inverse of synchrotron frequency.
Solving them, we get
\bea
\alpha&=&\left(\frac{-{\tilde e}E\tau_c}{\om}\right)\frac{1}{1+(\tau_c/\tau_B)^2}
\nn\\
\beta&=&({\hat e}\cdot{\hat h})(\tau_c/\tau_B)^2\alpha
=\left(\frac{-{\tilde e}E\tau_c}{\om}\right)({\hat e}\cdot{\hat h})\frac{(\tau_c/\tau_B)^2}{1+(\tau_c/\tau_B)^2}
\nn\\
\gamma&=&(-\tau_c/\tau_B)\alpha
=\left(\frac{{\tilde e}E\tau_c}{\om}\right)\frac{(\tau_c/\tau_B)}{1+(\tau_c/\tau_B)^2}~.
\label{phi_coeff}
\eea
In terms of these coefficients, now we can write
\be
\phi=\frac{e\tau_c}{1+(\tau_c/\tau_B)^2}\frac{k_i}{\om}\{\delta_{ij}
-(\tau_c/\tau_B)\ep_{ijk}h_k+(\tau_c/\tau_B)^2h_ih_j\}E^j~.
\label{phi_Ej}
\ee
Now, the connection between Eqs.~(\ref{macro_E}) and (\ref{micro_E}) becomes
\bea
\sigma^{ij}E_j&=&J^i_D
\nn\\
&=&g{\tilde e}\int \frac{d^3\vk}{(2\pi)^3}\frac{k^i}{\om}\delta f
\nn\\
&=&g{\tilde e}\beta\int \frac{d^3\vk}{(2\pi)^3}\frac{k^i}{\om}\phi f_0(1\mp f_0)~,
~[{\rm since}~ \delta f=-\phi\frac{\del f_0}{\del \om}=\phi\beta f_0(1\mp f_0)]
\nn\\
\Rightarrow \sigma^{ij}
&=&\delta^{ij}\sigma_0
-\ep^{ijk}h_k\sigma_1+h^ih^j\sigma_2~,
\eea
where
\be
\sigma_n = g{\tilde e}^2 \frac{\beta}{3}\int \frac{d^3\vk}{(2\pi)^3}\tau\frac{\vk^2}{\om^2}
\frac{\tau_c(\tau_c/\tau_B)^n}{1+(\tau_c/\tau_B)^2} f_0(1\mp f_0)~.
\label{cond_n}
\ee
If we consider ${\hat h}$ in z direction, then conductivity matrix elements will be
$\sigma^{xx}=\sigma^{yy}=\sigma_0$, $\sigma^{xy}=-\sigma^{yx}=-\sigma_1$, $\sigma^{zz}=\sigma_0 +\sigma_2$
and remaining are zero. We can used other notations $\sigma_{\parallel}=\sigma_{zz}$,
$\sigma_{\perp}=\sigma_{xx}$, $\sigma_{\times}=\sigma_{xy}$, which represents parallel, perpendicular and Hall component in more distinctly.

\end{document}